\documentclass[useAMS,usenatbib,usegraphicx]{mn2e}
 
\title[Galaxy Structure and Stellar Populations at $z > 1$]{The Tumultuous Formation of the Hubble Sequence at $z > 1$ Examined with HST/WFC3 Observations of the Hubble Ultra Deep Field}
\author[C.J. Conselice et al. ]{C. J. Conselice$^{1}$\thanks{E-mail: conselice@nottingham.ac.uk}, A.F.L. Bluck$^{1,2}$, S. Ravindranath$^{3}$, A. Mortlock$^{1}$, A. Koekemoer$^{4}$, \newauthor F. Buitrago$^{1}$, R. Gr\"utzbauch$^{1}$, S. J. Penny$^{1,5}$ \\
$^{1}$University of Nottingham, School of Physics \& Astronomy, Nottingham, NG7 2RD UK \\
$^{2}$Gemini Observatory, Hilo Hawaii 96720, USA \\
$^{3}$IUCAA, Pune University Campus, Pune 411007, Maharashtra, India \\
$^{4}$Space Telescope Science Institute, Baltimore, MD USA 21218 \\
$^{5}$Endeavour Fellow, Swinburne University of Technology, Hawthorn, Victoria 3122, Australia }
\def\deg{$^{\circ}\,$}
\def\solm{M$_{\odot}\,$}

\def\deg{$^{\circ}\,$}
\def\solm{M$_{\odot}\,$}

\def\mass{$10^{11}$ M$_{\odot}\,$}

\def\casgm20{CAS-G-M$_{20}\,$}
\def\m20{M$_{20}\,$}
\begin{document}

\date{Accepted ; Received ; in original form}
\pagerange{\pageref{firstpage}--\pageref{lastpage}} \pubyear{2002}

\maketitle

\label{firstpage}

\begin{abstract}

We examine in this paper a stellar mass selected
sample of galaxies at $1 < z < 3$ within the Hubble Ultra Deep Field, utilising
WFC3 imaging to study the rest-frame optical morphological distribution of galaxies 
at this epoch.  We measure how apparent morphologies 
(disk, elliptical, peculiar) correlate with 
physical properties, such as quantitative structure and spectral-types.  One primary
result is that apparent morphology does not correlate strongly with stellar populations,
nor with galaxy structure at this epoch, suggesting a chaotic formation history for
Hubble types at $z > 1$.   By using a locally defined definition of disk 
and elliptical galaxies based on structure and spectral-type, we find no true ellipticals at 
$z > 2$, and a fraction of $3.2\pm2.3$\%
at $1.5 < z < 2$.  Local counterparts of disk galaxies are at a similar level of $7-10$\%, much lower
than the 75\% fraction at lower redshifts.
We further compare WFC3 images with the rest-frame UV view of galaxies from ACS
imaging, showing that galaxies imaged with ACS that appear peculiar often contain 
an `elliptical' like morphology in WFC3.   We  
show through several simulations that this larger fraction of 
elliptical-like galaxies is partially due to the courser PSF of WFC3, and that the
`elliptical' class very likely includes early-type disks.  We also measure the merger history 
for our sample using CAS parameters, finding a 
redshift evolution increasing with redshift, and a peak merger fraction of 
$\sim 30$\% at $z \sim 2$ for the most massive galaxies with $M_{*} > 10^{10}$ \solm,
consistent with previous results from ACS and NICMOS.  We also compare our results 
to semi-analytical model results and find a relatively good agreement between
our morphological break-down and the predictions.
We finally argue that galaxies classified visually as peculiar, 
elliptical and as peculiar ellipticals, all have similar structural and
stellar population properties, suggesting that these galaxies are 
in a similar formation mode, likely driven by major mergers.

\end{abstract}


\section{Introduction}

Explaining the origin of the Hubble sequence of galaxies has remained one of the most
outstanding problems in extragalactic astronomy.  Massive galaxies in the 
nearby universe are largely in the form of ellipticals and spirals. 
These galaxies tend to not be undergoing significant galaxy or
star formation processes,
and are the end result of evolution and formation over the past 
$\sim$ 13.7 Gyr.   However, deep Hubble Space Telescope (HST) imaging using both 
optical and near-infrared (NIR) cameras, such as ACS and NICMOS, have shown 
convincingly that the galaxy 
population at higher
redshifts, particularly at $z > 2$, is dominated by irregular and peculiar
galaxies that often have no obvious similarity, in terms of structure,
to lower redshift galaxies (e.g., Driver et al. 1995; Abraham et al. 1996;
Conselice et al. 2005, 2008; Lotz et al. 2004; Cassata et al. 2005, 2010; Overzier
et al. 2010; Cameron et al. 2010).  
In fact the structures and morphologies of galaxies, and how they evolve,
are now recognised as one of the most important methods for 
understanding galaxies, as it reveals not only the empirical evolution of
a new parameter, but furthermore correlates with physical processes, such as 
merging,
occurring within galaxies (e.g., Conselice 2003; Conselice et al. 2003b;
Cassata et al. 2005; Grogin et al. 2005;  Trujillo et al. 2007).

A major remaining issue in galaxy formation is solving the riddle of how the Hubble
sequence came to be, and when this formation occurred.
This problem of understanding the origin of the Hubble sequence can be defined 
most simply as uncovering 
when and how the disk and elliptical population in the nearby universe
developed.  This is a difficult problem since the population of disks and
ellipticals is largely in place by $z \sim 1$ (Conselice et al. 2005) and to
understand the formation of these systems one has to probe higher redshifts
where the NIR is needed to resolve the location of most of the stellar mass
in these galaxies.  

There are however some existing clues for how this formation occurs
from deep Hubble Space Telescope imaging, particularly from observations with the 
NICMOS camera (e.g.,
Conselice et al. 2005; Conselice et al. 2011).  These observations show that, 
at rest-frame optical
wavelengths, the types of galaxies seen changes gradually with time, such
that disks and ellipticals are the dominate population at $z < 1$, while
peculiars are the most common at $z > 2$ (Conselice et al. 2005, 2008).  
The population of `Hubble
sequence' galaxies roughly matches that of the peculiars sometime
between $z = 1.5 - 2$, which is also the 
peak of the star formation
rate in the universe (e.g., Bouwens et al. 2009).

There are at least two issues with these observations that still need
to be addressed, and where HST Wide-Field Camera-3 (WFC3) observations 
will make a big impact
in the next few years. One of these questions  is 
understanding when the transformation between peculiar galaxies into disks 
and ellipticals occurs. The
other major question is understanding the physical mechanisms responsible
for driving the formation of these Hubble sequence galaxies.  For
example, are ellipticals and spirals formed through the accretion of
gas, such as in cold streams (e.g., Dekel et al. 2009), and/or through 
in-situ formation of stars within an existing dark matter halo, and/or
through mergers of existing galaxies (Hopkins et al. 2006)?

Recent observations of the kinematic structures of high redshift galaxies
show that there is a large diversity of types, but with many star
forming galaxies having a large velocity dispersion compared to their
rotation (Genzel et al. 2006).  This is either an indication that these systems are
large disks collapsing, or that they are undergoing some type of
merging activity (Shapiro et al. 2008).  Overall, kinematic studies
of $z > 2$ galaxies find that up to
1/3 of these systems have merger signatures in
their kinematics, which is consistent with the fraction found from imaging
(Conselice et al. 2003, 2008).
However, these results are not entirely fool-proof, as other formation
modes can mimic both the kinematic and morphological signatures. Furthermore
the samples studied kinematically  remain quite small, at a few 10s of systems. 
The relative role of these various assembly processes at high-$z$ remain an
open and important question regarding the mechanisms by which galaxy formation
has occurred since $z \sim 3$.

New deep NIR observations of the high-redshift universe can provide some answers 
to these
questions by examining the first question of when the Hubble sequence was
in place, and to begin addressing the formation mechanisms for 
galaxies we see in today's universe.  Observations with
the WFC3 of the Hubble Ultra Deep Field provides our first opportunity
to go beyond NICMOS observations that currently exist.  We describe a preliminary investigation
of this problem using this early data and find that an apparently mostly smooth
`early-type' population
appears to be the dominate one at high redshift, as compared with late-type or disk
dominated galaxies. We further discuss the merger history
and how it appears that high redshifts $1 < z < 3$ were not a particularly hospitable time
for the formation of prominent late-type disks.

\begin{figure}
\hspace{-0.5cm}
 \vbox to 130mm{
\includegraphics[angle=0, width=90mm]{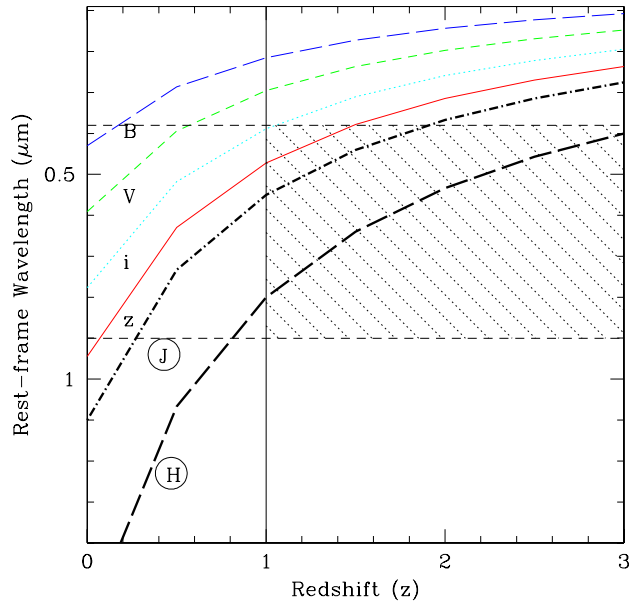}
 \caption{The rest-frame wavelength probed as a function of redshift
for the two filters used in this study - J$_{110}$ and
H$_{160}$ from WFC3 compared with the ACS filters of $B_{450}$, $V_{606}$,
$i_{775}$ and $z_{850}$.  The vertical line at $z = 1$ denotes the lower 
limit to redshifts  considered in this paper.  The shaded region shows
the region where the rest-frame optical is probed, showing that for all but
near $z \sim 1$ one must use NIR imaging to study the rest-frame structures
of these galaxies.}
%\vspace{3cm}
} \label{sample-figure}
\end{figure}

%fig uses massz from plot.sm and plotwf.sm
\begin{figure*}
%\vspace{5.5cm}
 \vbox to 115mm{
\includegraphics[angle=0, width=184mm]{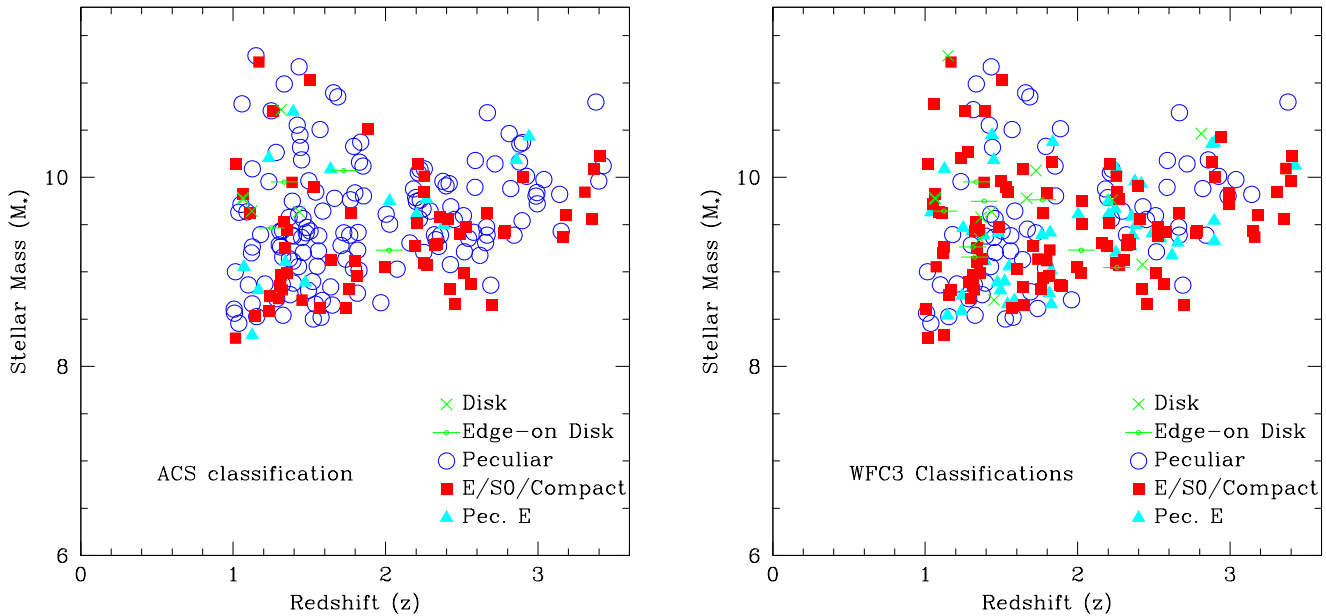}
 \caption{Stellar mass vs. redshift with the various galaxy
types studied in this paper labelled. On the left we show the
classifications for $z > 1$ galaxies carried out on ACS
imaging of the UDF, and the right panel shows the classifications
done using WFC3 at the rest-frame optical. The open blue circles are
the galaxies classified visually as peculiars, the solid red boxes are ellipticals,
S0s, and compacts, while the cyan triangles are ellipticals that appear to
have a peculiarity.  Disk galaxies are shown as green crosses for face
on systems, with the edge-on systems displayed as a dot with a solid
line.  }
\vspace{-3cm}
} \label{sample-figure}
\end{figure*}

This paper is organised as follows.  \S 2 gives an overview of the data and
data sources we use, including previous work which we use to analyse
our sample. \S 3 is a description of our structural measures, including CAS
and S{\'e}rsic profiles fits, as well as our measures of the spectral types of
our galaxies. \S 4 discusses our results, including morphological $k$-corrections,
the types of galaxies found at high redshifts and a comparison to models, 
and how various classification methods
compare with each other, while \S 5 is a summary of the results of this paper.
We use a standard cosmology of {\it H}$_{0} = 70$ km s$^{-1}$ Mpc$^{-1}$, and 
$\Omega_{\rm m} = 1 - \Omega_{\Lambda}$ = 0.3 throughout.

\section{Data and Data Sources}

The primary data used in this paper is the WFC3 
observations of the Hubble Ultra Deep field located in the
GOODS-South field (Giavalisco et al. 2004).  We also use ACS and NICMOS
imaging of the Hubble Ultra Deep Field (UDF)(Thompson et al. 2005;
Beckwith et al. 2006) within this analysis, particularly when
comparing with previous structural results (Conselice et al. 2008).  
The WFC3 field of view is 11 
arcmin$^{2}$, with a drizzled pixel scale of 0.03 arcsec per pixel, the same
as for the ACS imaging. The filters we use in this study
are the F160W (H$_{160}$) and the F110W (H$_{110}$) 
with a depth similar to the optical down to roughly $m_{\rm AB} \sim 29$
using 16 orbits in the J-band and
28 orbits in H.

The UDF ACS images are taken within the F435W (B$_{435}$), F606W 
(V$_{606}$), F775W ($i_{775}$), and F850L ($z_{815}$) bands.   The central
wavelengths of these filters, and their full-width at half-maximum
are: F435W (4297, 1038 \AA), F606W (5907, 2342 \AA), 
F775W (7764, 1528 \AA), F850L (9445, 1229 \AA).  In
Figure~1 we show the rest-frame wavelength probed by the WFC3 filters
we use, as well as those for the ACS, demonstrating the necessity of
using the near-IR to probe the rest-frame optical at $z > 1$ (shown as the
dashed area).  

The photometry and photometric redshifts we use are taken from our original
study of the Hubble Ultra Deep Field (Conselice et al. 2008).  The redshifts
we use are primarily from Coe et al. 
(2006) who measure the photometry of galaxies in the UDF 
within the BV$iz$JH bands.  The galaxies are detected 
with a modified version of SExtractor, called SExSeg. The
photometry is PSF-corrected and aperture matched, removing the
problem of matching magnitudes at different wavelengths due to 
variations in the PSF.   This high fidelity photometry is then used
to derive photometric redshifts, and for the stellar masses we measure
for our galaxy sample.
  
The Coe et al. (2006) photometric redshifts are measured using the photometric
redshift techniques from Benitez (2000), based on the optical and NIR 
photometry from ACS and NICMOS. In addition to these photometric 
redshifts we utilise 56 spectroscopic redshifts
within the UDF field within our selection limits.  All of the morphologies
we determine in the near-infrared are examined with WFC3, and all the optical
morphologies and structures with the ACS camera.

As this is a morphological, and not a photometric study, we must have
more stringent constraints on which galaxies we can examine.  We limit 
our study to systems which have magnitudes $z_{850} < 27$, such that we are 
not biased by low signal to noise imaging.     This final ACS based 
catalogue of $z_{850} < 27$ sources contains 
1052 unique galaxies within the ACS UDF field.  Of these 299 are found within
the smaller field of view WFC3 imaging of the UDF within the redshift
range $1 < z < 3$.   Note that although we use a $z_{850}$ limit to define our
sample, we later limit our analysis in the various analyses we carry out to
a subsample of these.
We do not include a discussion of lower or higher redshift galaxies in this paper.  
To run our morphological analysis requires segmentation maps, and we ran
SExtractor on the WFC3 imaging to obtain these maps which we later use within
the morphological analysis (see Conselice et al. 2008).

\section{Structural Measures, Stellar Masses, and Spectral Types}

In this paper we examine the  morphological, structural and spectral properties 
of $z_{850} < 27$ galaxies in the UDF, as imaged in the WFC3 UDF pointing.
We use an ACS $z_{850}$ selection as this was done in our previous analyses of
the UDF (Conselice et al. 2008).  This allows us to compare directly how
galaxy morphological and structural parameters vary with observed
wavelength.  We carry out several analyses of our WFC3 imaging using
structural parameters, as well as stellar masses, and
spectral types.  We carry out these analysis in an independent way,
thus lowering the effects of systematic errors on our overall results.

We carry out our visual and CAS analysis in several steps to maximise the 
usefulness
of the data, and to minimise problems from contamination.
We first create postage stamp images of each of our sample galaxies from
our WFC3 mosaic.
These are created by cutting out a 10'' $\times$ 10'' box of the UDF 
surrounding each galaxy, based on positions from the SExtractor
catalog detections from Coe et al. (2006).   Before this is done, the UDF
WFC3 cutouts are cleaned of projected nearby galaxies
and stars through the use of the  `segmentation map' produced
by SExtractor (see Conselice et al. 2008).   These segmentation maps
are equivalent in size to the UDF image itself, with the difference
being that it gives a numerical value for each pixel that reveals
which galaxy it belongs to.  These segmentation maps are used for
photometry, but they are also useful for removing nearby galaxies.
The procedure we use is to replace pixels of galaxies not being studied
to the sky background with proper noise characteristics included.  We then 
use these cleaned cutout images in our analysis.

%fig uses massz from plot.sm and plotwf.sm
\begin{figure}
%\vspace{5.5cm}
 \vbox to 120mm{
\includegraphics[angle=0, width=84mm]{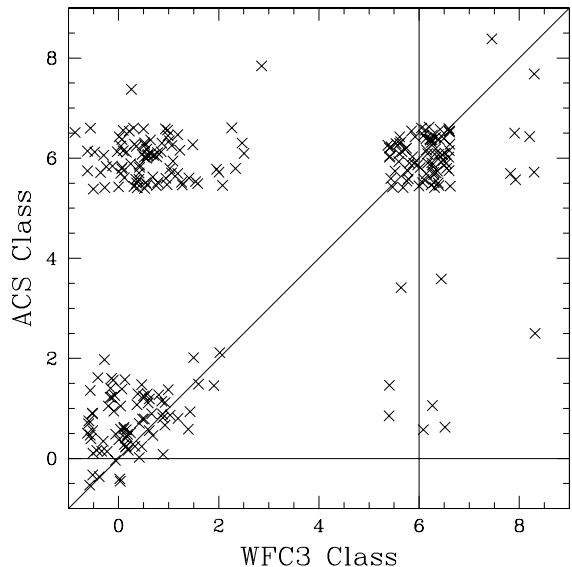}
 \caption{The relationship between the classification of the same
galaxies within ACS and WFC3. The following key corresponds to the
number schemes: 0 = Elliptical, 0.5 = peculiar elliptical, 1 = compact
elliptical; 2 = early-disk, 3=late-disk , 6 = peculiar, 8 = edge-on disk
galaxy.  Note that the points on this plot are randomly offset to show
the relative number at each position.}
\vspace{3cm}
} \label{sample-figure}
\end{figure}

After examining our sample visually, we found that occasionally features
remained near galaxies, and had to be manually removed by hand. There were 
also cases where large late-type galaxies with spiral arms 
brighter than their centres tended to be picked up by the program more than 
once, and
these were manually noted when spotted.  In the following
sections we describe our visual and quantitative analysis of these galaxy
images within the UDF.

\begin{figure*}
%\vspace{5.5cm}
 \vbox to 100mm{
\includegraphics[angle=0, width=184mm]{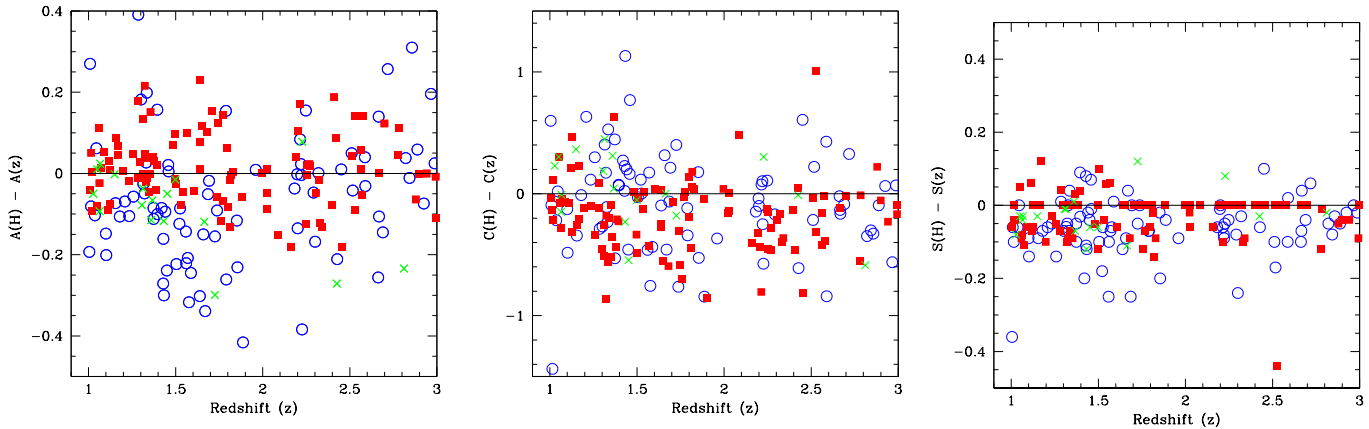}
%\vspace{-0cm}
 \caption{The change in the CAS  parameters
as a function of redshift. We trace in this figure the change
between the observed $\sim$1.6 $\mu$m ($H_{160}$) 
and $\sim$0.9 $\mu$m ($z_{850}$) morphology.  The symbol shape and
colour denotes the galaxy type it denotes, as in Figure~2.  The
open blue circles are the galaxies classified visually as peculiars,
the solid red squares are early-types, while the green crosses
are disk galaxies.  The horizontal line labels where there is
no change between the measure structure in these two bands. 
Compared to morphological changes at $z < 1$ between rest-frame
UV and optical, we find small changes at higher redshifts. }
} \label{sample-figure}
\end{figure*}

Likewise, we perform a parametric analysis on our data using the GALFIT
package (Peng et al. 2002) to determine parametric sizes and S{\'e}rsic
indices (\S 3.3).  We use these S{\'e}rsic indices as another indicator
of galaxy 'type' -- i.e., what are likely progenitors of modern day disks
and ellipticals.  Furthermore, we also use the spectral energy distributions
of our galaxies to determine their spectral types, which we
explain in \S 3.4.  By using spectral types we have a third and largely
independent way to determine galaxy types, or their progenitors, at higher
redshifts. In the case, instead of structure, we are examining their
stellar populations - namely, early-type, disk, and starburst/Im.

\subsection{Visual Morphologies}

We  use several
different methods in this paper to examine how the structures and morphologies
of galaxies change with redshifts.  First, one method involves a visual 
estimate of morphologies 
based on the appearance of our galaxies in the WFC3 imaging.   The basic
outline of how we carry out 
our classification process is given in Conselice et al. (2005a),
Conselice et al. (2007b) and Conselice et al. (2008).    
The process is that we place each galaxy in the WFC3 imaging into one of nine 
categories: compact, elliptical, distorted elliptical, lenticular (S0),
early-type disk,  late-type disk, edge-on disk, merger/peculiar, 
and unknown/too-faint.  These classifications are based only 
on how a galaxy looks, as seen in the observed H$_{160}$-band within
WFC3.   These galaxies are all resolved within at least 3-4 times the
effective radius.   We do not use information such as colour, size, redshift, etc 
to determine our classification types.  Note that with WFC3, the resolution is
not generally very high for these distant galaxies, and often galaxies
look like 'blobs', and a classification type is sometimes more of an educated guess, 
and certainly prone to variation between classifiers. However, our classifications were done
consistently, as described below.  First we give the main types in which
our classifications were done.

\begin{enumerate}

 \item Ellipticals : Ellipticals (Es) are centrally concentrated objects with 
no evidence for lower surface brightness, outer structures.  We have
86 of these galaxies in our sample.  In this paper, the meaning of
`elliptical' or `early-type' henceforth is likely not the same as a pure
elliptical selection at low redshift.  Simulations (\S 4.3) show
that what we classify as `elliptical' is likely a mixture of true ellipticals
and spiral galaxies dominated by a bulge and/or mergers in a more relaxed
phase.

 \item Peculiar-Ellipticals : Peculiar ellipticals  (Pec-Es) are galaxies that
appear elliptical, but have some minor morphological peculiarity,
such as offset isophotes, dual nuclei, or low-surface brightness asymmetries
in their outer parts (58 systems). A full description of these galaxies
is provided in Conselice et al. (2007).

  \item S0s: S0s are galaxies that appear to have a smooth disk with
a bulge. These galaxies do not appear to have much star formation,
and are selected in the same way nearby S0s are.  Our sample
has only a single system making its contribution very small.

 \item Compact - A galaxy is classified as compact if its structure
is resolved, but still appears compact without any substructure. It is 
similar to the elliptical classification in that
a system must appear very smooth and symmetric to be included. 
A compact galaxy differs from an elliptical in that it contains no 
features such as an extended
light distribution or a light envelope.  The fact that these systems
are common at $z > 1$ while ellipticals are more commonly seen at
lower redshifts suggests that they are potentially drawn from the
same population.  (28 systems)
     
 \item Early-type disks: If a galaxy contains a central 
concentration with some evidence for lower surface brightness outer 
light in the form of spiral arms or a disk, it is classified as an 
early-type disk. (14 systems)
 
 \item Late-type disks: Late-type disks are galaxies that appear to 
have more outer low surface brightness disk light than inner concentrated 
light. (4 systems)

 \item Edge-on disks: disk systems seen edge-on, and whose face-on morphology 
cannot be determined, but is presumably an S0 or spiral. (9 systems)
       
 \item Peculiar/irregular: Peculiars and irregulars are systems that 
appear to be disturbed, or peculiar looking, including elongated/tailed 
sources. These galaxies are possibly in some phase of a merger 
(Conselice et al. 2003a), or are dominated by star formation (89 systems).
   
 \item Unknown/too-faint: If a galaxy is too faint for any reliable 
classification it was placed in this category. Often these galaxies appear 
as smudges without any structure. These could be disks or ellipticals, but 
their extreme faintness precludes a reliable classification.   We find
that there are no systems in this category using our selection, although
unclassifiable systems are found in the UDF ACS imaging.

Furthermore, we find that nine of our galaxies are not classifiable as they
are unresolved, appearing as point sources. This may be misidentified stars,
QSOs, or a mixture of these. They are not included in further discussions 
in this paper.  Furthermore, we performed these same classifications on the
ACS $z_{850}$-band imaging of the UDF, in the same manner as for the H$_{160}$
imaging, for comparison purposes.

%fig uses plotwf.sm hmorph hmorphm
\begin{figure*}
%\vspace{5.5cm}
 \vbox to 90mm{
\includegraphics[angle=0, width=184mm]{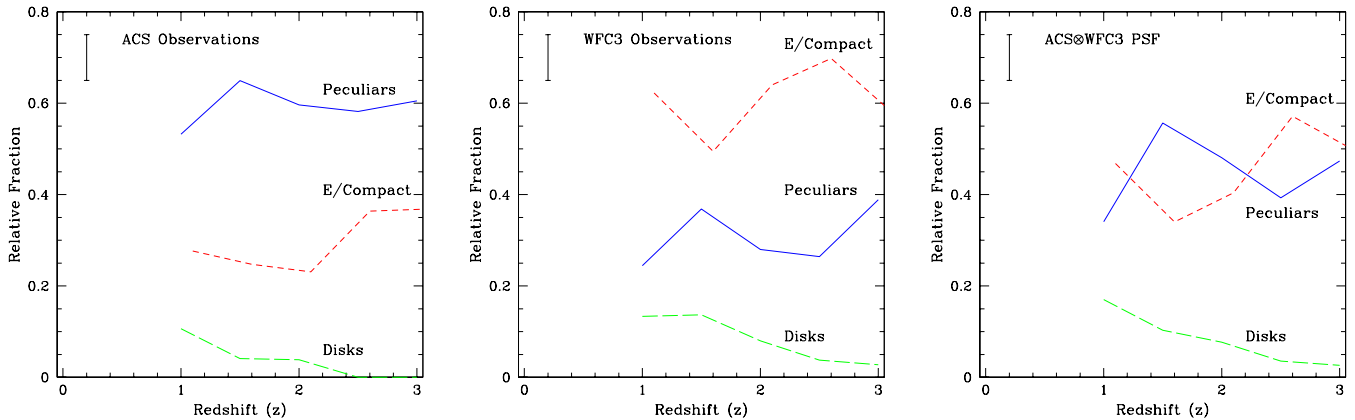}
 \caption{The relative distribution of galaxy types in the WFC3 Hubble
Ultra Deep Field for systems selected with $z_{850} < 27$ as measured
within three different images.  The left panel shows the distribution of
our classifications using the ACS imaging of the UDF (Conselice
et al. 2008). The middle panel shows our new WFC3 UDF classifications
and the right panel shows these same classifications done on 
a WFC3 image convolved with the ACS point spread function.   Labelled
on each panel are disks, ellipticals/S0s, compact galaxies, and peculiars. Note
that the compact galaxies become an  important population at higher
redshifts.  }
\vspace{-3cm}
} \label{sample-figure}
\end{figure*}

\end{enumerate}

\subsection{CAS Parameters}

In addition to visual morphologies, we utilise the CAS 
(concentration, asymmetry, clumpiness) parameters to probe the 
structures of our galaxies quantitatively.  The CAS parameters are a 
non-parametric method for measuring the forms of galaxies on resolved 
CCD images (e.g., Conselice et al. 2000a; Bershady et al. 2000; 
Conselice et al. 2002;  Conselice 2003).   The
basic idea behind the CAS system (Conselice 2003) is that galaxies have 
light distributions revealing their past and present formation modes. 
This system can also be used to find specific galaxy types, as defined
in the nearby universe, such as ellipticals and mergers/peculiars.  For
example,  the 
selection $A > 0.35$ locates systems which are nearly all major
galaxy mergers in the nearby universe (e.g., Conselice et al. 2000b; 
Conselice 2003; Hernandez-Toledo et al. 2005; Conselice 2006b).

The basic
procedure for measuring the CAS parameters is essentially the same as
that presented in Conselice et al. (2008).  First the galaxy is
cut out from the main UDF WFC3 image from which the radii and the
CAS parameters are measured.  We measure the Petrosian radii for each
galaxy from our images. This is the radius 
defined as the location where the surface brightness at a given
radius is 20\% of the surface brightness within that radius (e.g.,
Bershady et al. 2000; Conselice 2003).    The exact
Petrosian radius we use to measure our parameters is

$$R_{\rm Petr} = 1.5 \times r(\eta = 0.2),$$

\noindent where $r(\eta = 0.2)$ is the radius where the surface
brightness is 20\% of the surface brightness within that radius.
 
We use circular apertures to measure
our Petrosian radii and quantitative parameter estimation.  We begin
our estimates of the galaxy centre for the radius measurement at
the centroid of the galaxy's light distribution.   Based on various
test and simulations, we have found that the resulting radii do not
depend critically on the exact centre, although for other
parameters this can be more important 
(Conselice et al. 2000; Lotz et al. 2004).
For more detail see Bershady et al. (2000),
Conselice et al. (2000), Conselice (2003) and Conselice et al. (2008).

\subsubsection{Asymmetry Parameter, $A$}

The basic method for measuring the asymmetry is by taking an original 
galaxy  image and rotating it 180 degrees about its centre, and then
subtracting these two images (Conselice 1997).  Important corrections are 
done for background light, and the radius has to be measured carefully,
as explained in detail in Conselice et al. (2000a, 2002).  Details such as how
centring is done is also described in Conselice et al. (2000a).  The mathematical
definition is

\begin{equation}
A = {\rm min} \left(\frac{\Sigma|I_{0}-I_{180}|}{\Sigma|I_{0}|}\right) - {\rm min} \left(\frac{\Sigma|B_{0}-B_{180}|}{\Sigma|I_{0}|}\right),
\end{equation}

\noindent where $I_{0}$ is the original image pixels, $I_{180}$ is the image
after rotating by 180\deg.  The background subtraction using light from a
blank sky area, called $B_{0}$, are critical for this process, and must 
be minimised in the same way as the original galaxy itself.  Lower values 
of $A$ imply that a galaxy is largely symmetric, which tends to be found in
early type galaxies.
Higher values of $A$ indicate an asymmetric light distribution, which are 
usually found in spiral galaxies,  or in the more extreme case, merger 
candidates. 

\subsubsection{Light Concentration $C$}

The concentration index ($C$) is a measure of the intensity of 
light contained within a central region defined by a curve of growth
radii, compared to a larger  curve of growth radii. The exact 
definition is the log of the ratio of two circular radii which 
contain 20\% and 80\% ($r_{20}$, $r_{80}$) of the total galaxy flux,

\begin{equation}
C = 5 \times {\rm log} \left(\frac{r_{80}}{r_{20}}\right).
\end{equation}

\noindent This index is sometimes called C$_{28}$.  
A higher value of $C$ indicates that a larger amount of light 
in a galaxy is contained within a central region.   Our measurement of the 
concentration correlates well with 
the mass and halo properties of galaxies (e.g., Bershady et al. 2000; 
Conselice 2003).   Nearby galaxy values for the concentration
index are $C = 2-3$ for disks, $C > 3.5$ for massive ellipticals, while
peculiars span the entire range (Conselice 2003).

\subsubsection{Clumpiness Parameter, $S$}

The clumpiness ($S$) is a parameter for describing 
the fraction of light in a galaxy contained in clumpy light
concentrations.   Galaxies that appear `clumpy', such as
star forming systems, have a relatively large amount of
light at high spatial frequencies, and high $S$ indices. 
Smoother galaxies, such as ellipticals contain smaller $S$
values.    Clumpiness can be 
measured in a number of ways, the most common method used,
as described in Conselice (2003) is,

\begin{equation}
S = 10 \times \left[\left(\frac{\Sigma (I_{x,y}-I^{\sigma}_{x,y})}{\Sigma I_{x,y} }\right) - \left(\frac{\Sigma (B_{x,y}-B^{\sigma}_{x,y})}{\Sigma I_{x,y}}\right) \right],
\end{equation}

\noindent where, the original image $I_{x,y}$ is blurred to produce 
a secondary image,  $I^{\sigma}_{x,y}$.  This blurred image is
then subtracted from the original image leaving a 
residual map, containing only high frequency structures in
the galaxy (Conselice 2003). To quantify this, we normalise the
summation of these residuals by the original galaxy's total light, and
subtract from this the residual amount of sky after smoothing
and subtracting it in the same way.  The size of the smoothing kernel 
$\sigma$ is
determined by the radius of the galaxy, and is $\sigma = 0.2 \cdot 1.5
\times r(\eta = 0.2)$ (Conselice 2003).  Note that the centres of galaxies are
removed when this procedure is carried out, typically the
central 1/8th of the Petrosian radius.

\subsection{Parametric Fitting -- S{\'e}rsic Indices}

While the non-parametric measures are one way to examine and quantify
galaxy structure, utilising parametrised fits to light distributions of galaxies
through an analytical form is another popular method for understanding
the structural forms of galaxies.  We fit to our WFC3 H$_{160}$-band
imaging a single S{\'e}rsic profile of the form:

\begin{equation}
\Sigma(r) = \Sigma_{e} {\rm exp}(-\kappa[(r/r_{e})^{1/n} -1])
\end{equation}

\begin{figure}
\hspace{-0.5cm}
 \vbox to 120mm{
\includegraphics[angle=0, width=90mm]{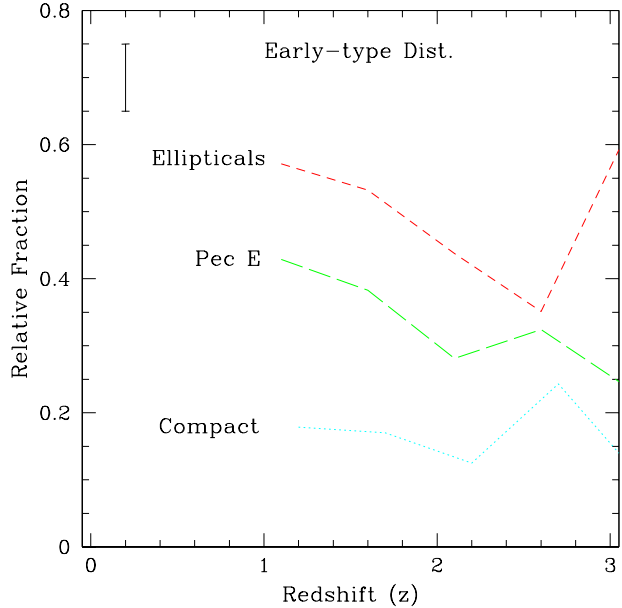}
 \caption{The distribution of galaxy types within the elliptical
category in Figure~5 divided into normal ellipticals, compact
ellipticals and peculiar ellipticals.}
\vspace{2cm}
} \label{sample-figure}
\end{figure}

\noindent We use the GALFIT package to fit equation (4) above to determine
the S{\'e}rsic index  ($n$) and the effective radius, (R$_{\rm e}$).  
GALFIT was used to generate 2-D models of the light profiles of our
sample galaxies using a single S{\'e}rsic function. The position, initial guess
parameters for GALFIT (total mag, R$_{\rm e}$, ellipticity, position angle,
etc.) were taken from the SExtractor catalog. The background value is
kept fixed at the value determined by SExtractor in the area around 
each galaxy. The constraints used when performing the
fitting includes requiring that the centroid remains within 3 pixels of the input 
position, and the S{\'e}rsic index, $n$, does not exceed a value of $n=10$. The PSF 
for convolving the analytical model was generated from the stars in UDF 
field using the PSF tasks in IRAF/DAOPHOT package, and then normalising these. 
When there are neighboring objects surrounding a specific object of interest, we
perform simultaneous fits to the neighbours, which has been found to
work better than masking the objects based on some isophotal threshold.  In this
paper we mostly make use of the S{\'e}rsic index, $n$, as a measure of the
`disk-like' or `elliptical-like' nature of the galaxies within our sample.  Typically
galaxies with fitted $n > 2$ are elliptical like, and $n < 2$ are disk-like in 
a traditional interpretation compared with nearby galaxies (e.g., Trujillo et al. 2007;
Buitrago et al. 2008).  With GALFIT we also measure the axis ratio of each system, giving
an ellipticity of $\epsilon = 1 - b/a$.

While it remains
to be determined whether the kinematic interpretation of objects with
rest-frame morphologies of a `disk' and `spheroid' nature match, the S{\'e}rsic
index is a good rough guide to the overall shape of the light distribution
within galaxies at different redshifts.  These fits were also done using
WFC3 data in Cassata et al. (2010) to investigate the size evolution of
the most massive galaxies at $z < 3$, finding that they are indeed compact,
as has been found in previous studies.  

\begin{figure*}
%\vspace{5.5cm}
 \vbox to 120mm{
\includegraphics[angle=0, width=184mm]{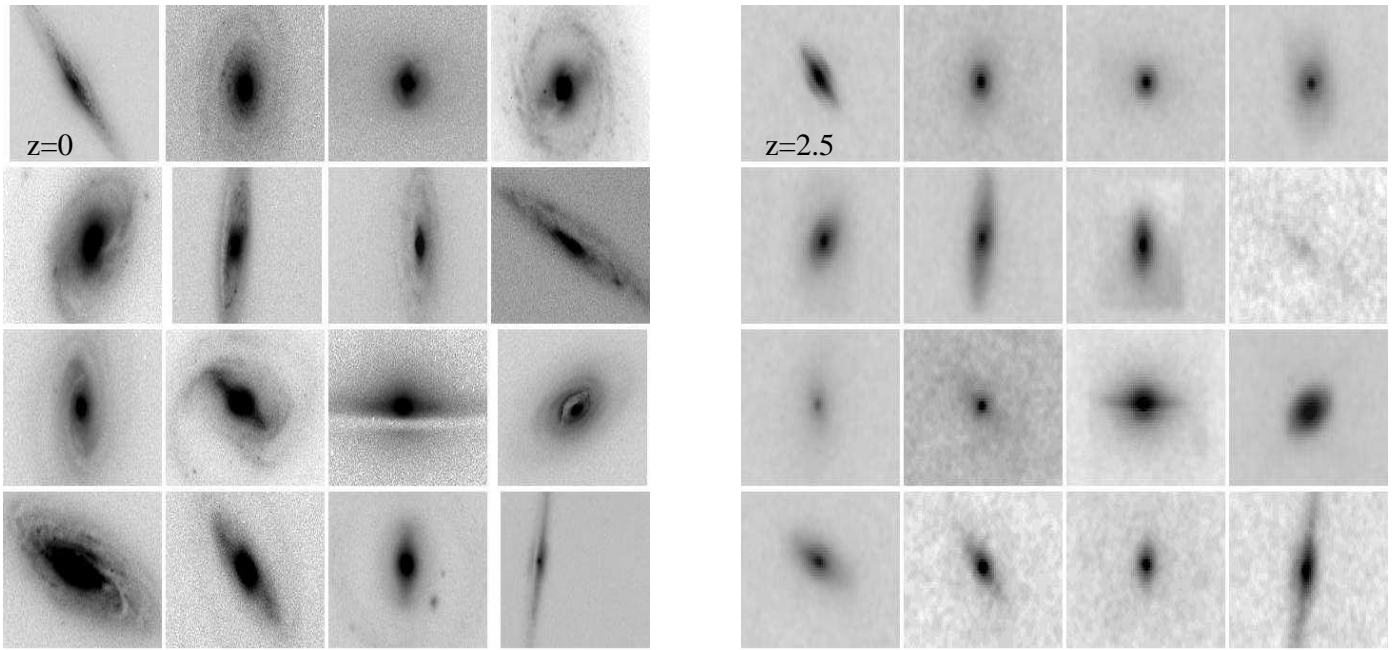}
\vspace{0cm}
 \caption{Example simulations of nearby galaxies which are originally observed
at $z \sim 0$ imaged to how they would appear at $z = 2.5$ within the Hubble Ultra Deep Field F160W band as imaged with WFC3. 
These nearby galaxies are classified as early-type spirals 
(Sa and Sb).  The typical sizes of these galaxies are several kpc in effective radii, 
and are at a variety of distances (see Conselice et al. 2000a).  As can be seen, many of 
these systems would be difficult to classify as early-type disks, or even as galaxies 
which contain a disk in some cases.}
} \label{sample-figure}
\end{figure*}
%\vspace{4cm}

\begin{figure*}
%\vspace{5.5cm}
 \vbox to 120mm{
\includegraphics[angle=0, width=184mm]{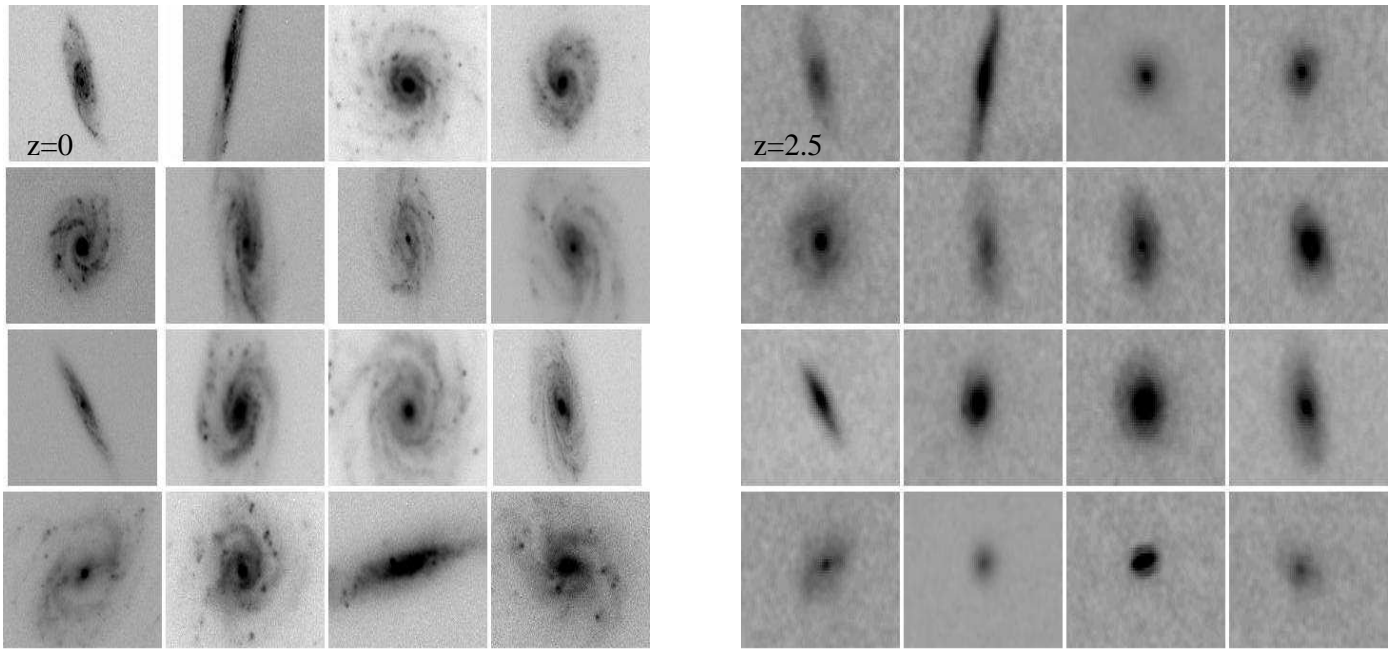}
%\vspace{-0cm}
 \caption{Example simulations of nearby galaxies which are originally observed
at $z \sim 0$ imaged to how they would appear at $z = 2.5$ within the Hubble Ultra Deep Field F160W band as imaged with WFC3. 
These nearby galaxies are classified as 
late-type spirals (Sc and
Sd).  As can been seen, several systems are edge on in these simulations.   
The typical sizes of these galaxies are several kpc in effective radii, and are at a variety
of distances (see Conselice et al. 2000a). }
} \label{sample-figure}
\end{figure*}
%\vspace{4cm}

\begin{figure*}
%\vspace{5.5cm}
 \vbox to 200mm{
\includegraphics[angle=0, width=164mm]{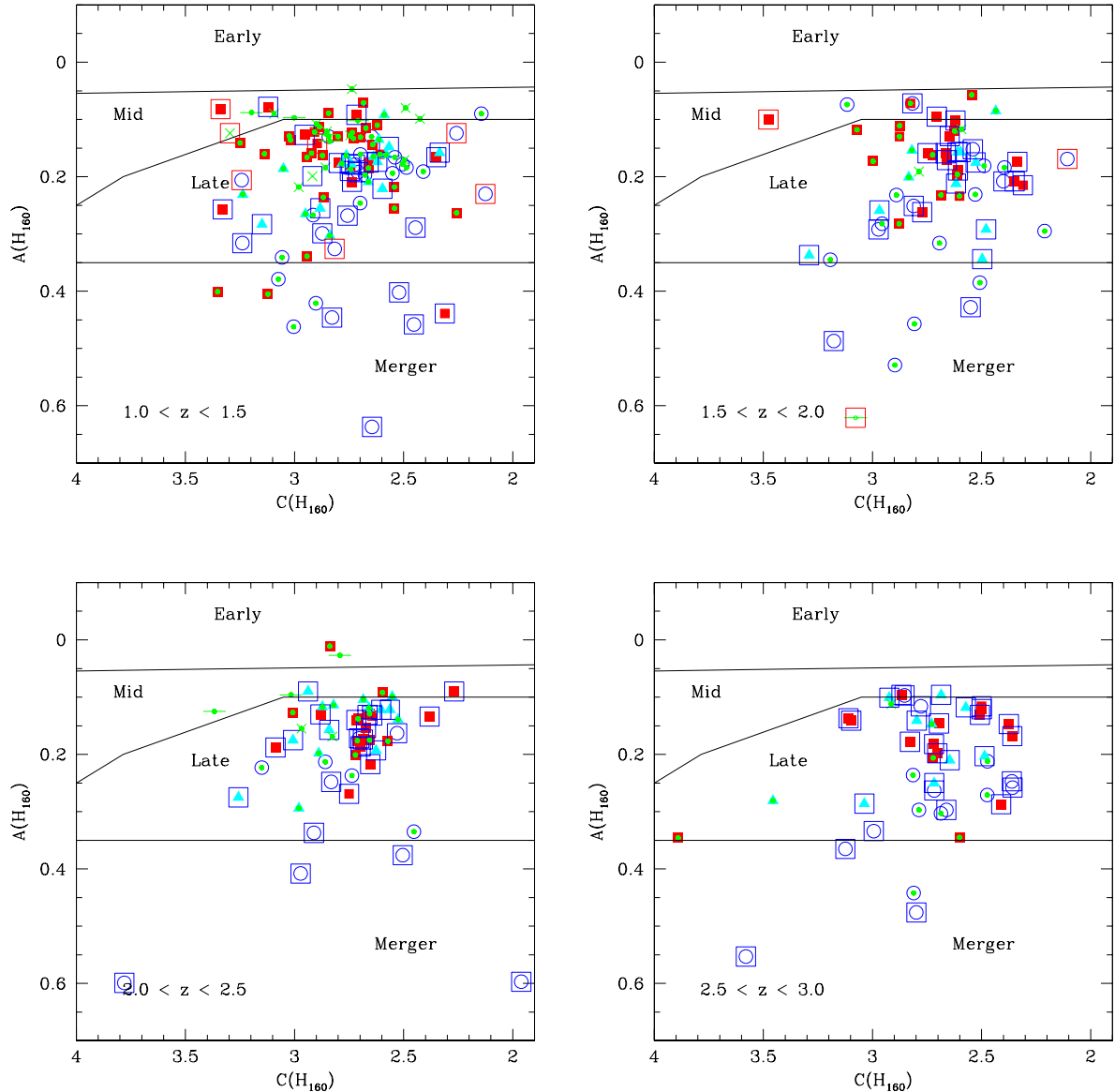}
%\vspace{-0cm}
 \caption{Plot showing the distribution of our WFC3 galaxy sample in the 
UDF as seen in the concentration-asymmetry plane of CAS space.  The lines on each plot
denote the region in which different galaxy types, as seen in the
rest-frame optical are found in the nearby universe (e.g., Bershady et 
al. 2000; Conselice 2003).  The redshift range for each panel is shown
and goes from $1 < z < 3$.  We show both the visual and spectral type classifications
within this figure.
The main visual symbol types for the points are the same as in Figure~1, such
that the open blue circles are the peculiars, the red boxes are the early-types,
the cyan triangles are the peculiar ellipticals, the green lines are edge-on
disks and the green crosses are disks/spirals.  We also show the spectral type 
classifications for the same galaxies in the following way: if the galaxy
has a red open box surrounding it, it is classified as a  early-type spectrally,
open blue boxes are starforming systems, Im/starbursts, and those that have a
green centre are classified spectrally as disk like or Sbc/Scd.}
} \label{sample-figure}
\end{figure*}
%\vspace{4cm}

\begin{figure*}
%\vspace{5.5cm}
 \vbox to 110mm{
\includegraphics[angle=0, width=164mm]{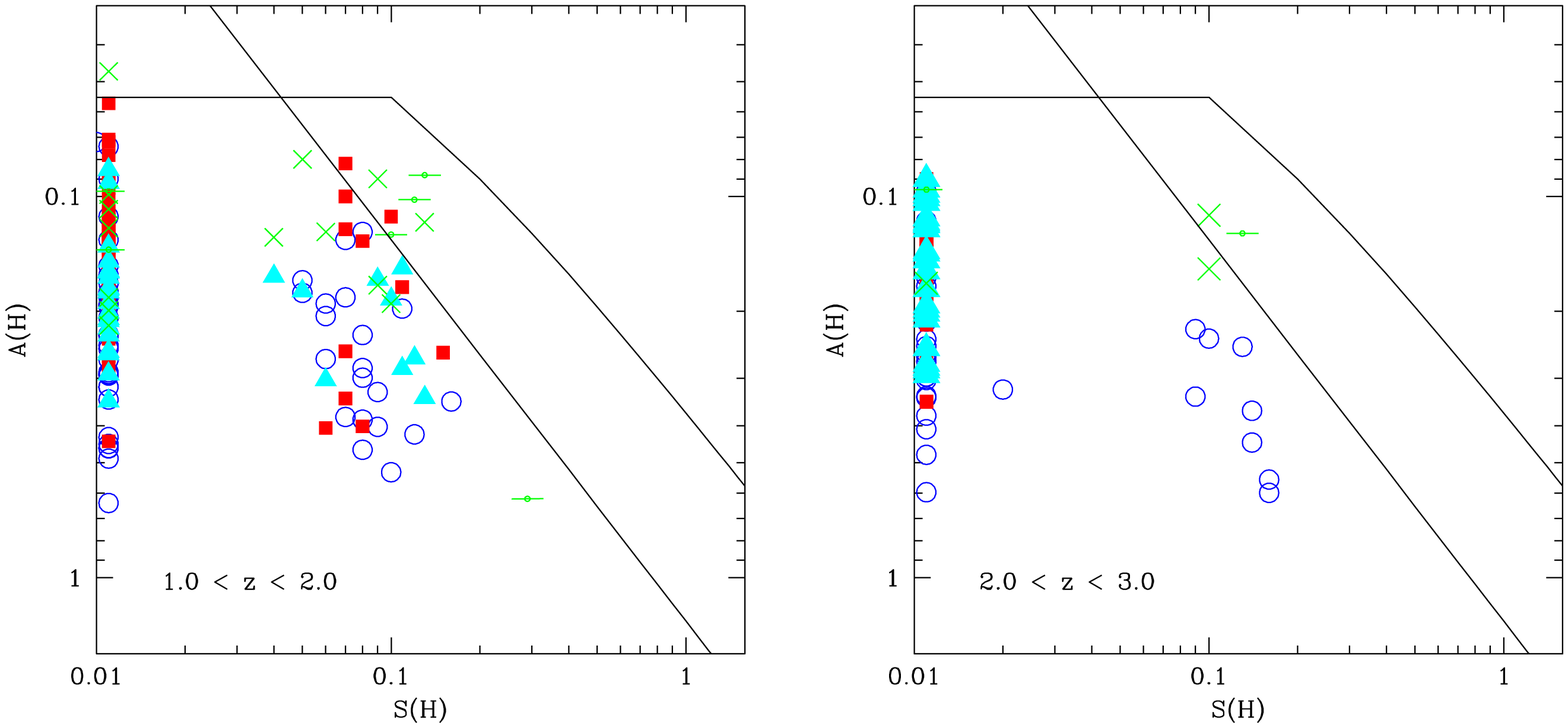}
%\vspace{-0cm}
 \caption{The relation between the asymmetry ($A$) and clumpiness ($S$)
for our sample of WFC3 UDF galaxies in the H$_{160}$ band.    
The solid line with the bend in it shows the relationship
between these two parameters as found for nearby galaxies that are not
involved in mergers (Conselice 2003). The straight line show the relationship
between these two parameters found at similar rest-frame wavelengths by
Lanyon-Foster et al. (2011).   This relationship is such that
non-merging galaxies with a higher clumpiness has a slightly higher
asymmetry, both due to star formation. Merging galaxies, where 
the structure is distorted due to bulk asymmetries from a merger
have a larger asymmetry for their clumpiness.  The point symbols
are the same as in Figure~1 and the redshift range for each individual
panel is listed. Note that galaxies which are generally more clumpy than
asymmetric are classified as disk/spirals visually.}
} \label{sample-figure}
\end{figure*}
%\vspace{4cm}

\subsection{Spectral Types}

Each galaxy in our sample has a morphological type which can be determined
by the CAS parameters, by visual estimates of morphology, or by the measured
S{\'e}rsic index.  We furthermore utilise in this paper another method of 
determining galaxy types through the relative ages of their luminosity
weighted stellar
populations through fitting model spectra to observed
spectral energy distributions (SEDs).

These observed spectral energy distributions are taken from the wealth
of ancillary photometric data in the UDF field.  In summary we use TFITed
magnitudes, including the ACS optical, Spitzer IRAC data, U-band ground
based photometry and deep IR photometry from ground-based telescopes.  In detail
these spectral types are based on imaging data from the VLT/VIMOS ($U$-band), 
HST/ACS imaging (F435W, F606W, F775W, and F850LP bands), VLT/ISAAC (J, H, and Ks bands), 
an Spitzer/IRAC bands at 3.6, 4.5, 5.8, and 8.0$\mu$m.  The spectral types used in these
fits include the SEDs based on the Coleman et al. (1980) types: E, Sbc, Scd, and Im based
on spectra from nearby galaxies.  In addition, starburst templates from Kinney 
et al. (1996) are included in these fits.  We utilise the spectral types from these fits to
compare with our morphological types.
Further details of the fitting code, the templates used can be found in Dahlen et al. (2010).

\section{Results}

In this section we analyse the morphological properties of our sample galaxies 
at $z > 1$.  These galaxies constitute the brightest systems as seen in the 
Hubble Ultra Deep field and are amongst the most massive galaxies at high 
redshifts.    We examine our sample in several ways.  One is to simply
look at the distribution of galaxy types at $z > 1$ which we have classified 
visually. The other is to examine how the morphologies of our systems differ 
between the WFC3 imaging and that seen in the ACS imaging of the same sample 
as examined in Conselice et al. (2008).  We also examine how 
morphology, as judged visually, compares with spectral-type classifications from 
SEDs.   Finally, we examine the structural parameters of our systems and 
compare what we calculate for the CAS values and S{\'e}rsic fits for our galaxies
and the spectral-types and visual morphological estimates for these systems.

\subsection{Morphological $k$-corrections}

In Conselice et al. (2008) we investigate the morphological $k$-correction for galaxies seen
in the UDF ACS field, particularly those at $z < 1$.  We carried this out
to determine how galaxies look different in a quantitative way at different
wavelengths, and then used a quantitative version of this correction to
put higher redshift galaxies at $z > 1$ at a pseudo rest-frame optical
wavelength from which their merger fraction was derived (Conselice
et al. 2008).  In this paper, we are now able to derive what is the
actual morphological $k$-correction between the $H_{160}$-band and the
$z_{850}$-band for $z > 1$ systems, or rather how rest-frame optical and rest-frame
ultraviolet morphologies differ.

First, to put this problem in context, the fundamental
issue with using ACS data to examine the morphologies of $z > 1$ galaxies is shown
in Figure~1 where the rest-frame wavelength probed by different ACS filters,
and our WFC3 filters, are shown.  This demonstrates that to probe the rest-frame
optical light of galaxies at $z > 1.3$ we must use the observed infrared, and that
the F160W (H$_{160}$) filter provides rest-frame optical coverage for galaxies
between $1 < z < 3$.  In the rest-frame optical, structural parameters such as
the CAS parameters are largely similar (Taylor-Mager et al. 2007; Lanyon-Foster et al.
2011).  Henceforth, later in this paper, we utilize mostly the H$_{160}$ imaging to probe 
the morphologies and structures of our galaxy sample.

There are several ways to compare morphological $k$-corrections and how they
evolve at $1 < z < 3$.  One way to do this is to compare estimates
of morphology in the observed $z_{850}$ ACS band with the WFC3 H$_{160}$ band. This
comparison is shown in Figure~3.    In general, galaxies along the 1:1 line have
the same classification, and those which differ significantly, further away from
this line, have increasingly different classifications in ACS and WFC3 

A few outstanding deviations should be noted on Figure~3. The first is that for 
the peculiar
class, as found in the ACS imaging, roughly half (78 out of 174) are also
classified as peculiars in the WFC3 imaging.  The remainder are disks (15 out of 174, 
or 9\%), or early-types/compact galaxies (73 out of 174, or 42\%).  This implies
that either roughly half of the peculiars seen at $1 < z < 3$ in the rest-frame
UV are actually elliptical-like in the rest-frame optical, or that the larger PSF of
the WFC3 imaging is creating galaxies that look smooth (see \S 4.2.1).  We
conclude that a significant fraction, but not all, of this change in type
between wavelengths is due to the larger PSF of WFC3 compared with ACS 
\S 4.3.

Otherwise, for those galaxies classified as early-type in the ACS image, we find
that almost all, or 75 out of 83 systems are also classified as an elliptical
of some form in the WFC3 imaging. The remaining systems
are mostly peculiars as seen by WFC3.   The other slight difference
in the classification between ACS and WFC3 is for systems classified
as peculiar within the WFC3 imaging. For these systems, we find that 78
out of 89 systems (88\%) are also classified as peculiar in the ACS imaging. 
The remainder are found to be early-types and two disks in the ACS imaging.

We furthermore investigate the quantitative morphological $k$-corrections for 
$z > 1$ galaxies
by comparing how the quantitative morphologies in the ACS $z_{850}$-band compares
with the same morphologies in the $H_{160}$-band.  Figure~4  shows the difference
in the CAS parameters for our sample of galaxies from $z \sim 1-3$.  What
we find is generally similar to the trends seen for similar rest-frame
wavelength morphological differences probed for $z < 1$ galaxies in Conselice
et al. (2008) (see Table~1 and compare with Table~1 in Conselice et al.
2008). That is, the morphological 
$k$-correction, particularly
for galaxies near $z \sim 1$ in the rest-frame optical and rest-frame
UV are fairly similar to what is found at $z > 1$.  

\vspace{1cm}
\setcounter{table}{0}
\begin{table}
 \caption{The change in CAS parameters from the rest-frame
optical to near-UV as function of redshift.  These values are defined
such that $\Delta = \lambda(H_{160}) - \lambda (z_{850})$}
 \label{tab1}
 \begin{tabular}{@{}lr}
  \hline
\hline
Peculiars & $z = 1 - 3$  \\
\hline
$\frac{\Delta C}{\Delta \lambda} (\mu m^{-1})$ & -0.59$\pm$2.10 \\ 
$\frac{\Delta A}{\Delta \lambda} (\mu m^{-1})$ & -0.21$\pm$0.68 \\
$\frac{\Delta S}{\Delta \lambda} (\mu m^{-1})$ & -0.24$\pm$0.30 \\
\hline
Ellipticals & $z = 1 - 3$  \\
\hline
$\frac{\Delta C}{\Delta \lambda} (\mu m^{-1})$ & -0.75$\pm$1.30  \\
$\frac{\Delta A}{\Delta \lambda} (\mu m^{-1})$ &  0.08$\pm$0.54 \\
$\frac{\Delta S}{\Delta \lambda} (\mu m^{-1})$ &  0.00$\pm$1.19 \\
\hline
Spirals & $z = 1 - 3$  \\
\hline
$\frac{\Delta C}{\Delta \lambda} (\mu m^{-1})$ &  0.01$\pm$1.20  \\
$\frac{\Delta A}{\Delta \lambda} (\mu m^{-1})$ & -0.34$\pm$0.48  \\
$\frac{\Delta S}{\Delta \lambda} (\mu m^{-1})$ & -0.09$\pm$0.23 \\
\hline

 \end{tabular}
\end{table}

As expected galaxies
are generally more asymmetric and clumpy when viewed at shorter wavelengths
compared to the H$_{160}$-band, and surprisingly less concentrated for the
ellipticals and peculiars.  Table~1 lists the quantitative morphological
$k$-correction for our galaxies between $z = 1$ and $z = 3$, using the same
quantitative method as in Conselice et al. (2008).  These values are the
difference in the measured H$_{160}$ morphology and the ACS
$z_{850}$ structural features.  

We discuss below each of the CAS parameters and how their values differ between
the H$_{160}$-band and the $z_{850}$-band.  
First, what we find from our measurements is that the concentration values are most
different between the $H_{160}$ and $z_{850}$ bands for the ellipticals, in the sense that the
ellipticals are more concentrated in the z$_{850}$-band than in the H$_{160}$-band. 
This is different from what is found for nearby galaxies (e.g.,
Taylor-Mager et al. 2007) where early types become more concentrated
at longer wavelengths.  We also find this in the ACS UDF imaging at
the lower redshifts ($z = 0.25 - 0.75$) where we find a positive
value of $\Delta C/\Delta \lambda = 0.43 \pm 1.5$.  However, we also
find that early-type galaxies, as classified visually, are less concentrated
in terms of their stellar mass distribution than in $z_{850}$-band light
(Lanyon-Foster et al. 2011).   This is likely due to central star formation dominating
the morphologies of these systems, even if they are classified as early-type.

We also find that the peculiar galaxies
within $z = 1 - 3$ have a similarly lower concentration as do ellipticals at longer
wavelengths (Figure~2; Table~1).  Spiral galaxies however, have
a concentration difference near zero, and on average there is no morphological
$k$-correction for these systems, likely due to the fact that these
spirals are dominated by star forming regions that are easily visible in
both the rest-frame UV and rest-frame optical.

In terms of the asymmetry parameter, we find a slight change with respect
to wavelength, which is less pronounced than what was found in Conselice
et al. (2008) for systems at $z = 0.75 - 1.25$, comparing
rest-frame 0.3 $\mu$m to 0.5 $\mu$m.  As would be expected, the peculiar galaxies  
and the spiral galaxies are more asymmetric on average in the $z_{850}$-band than in the
$H_{160}$-band at $1 < z < 3$ (Table~1).  However, we find that the elliptical
galaxies are slightly more asymmetric in the $H_{160}$-band on average, although
the difference is not large.

The clumpiness index has a similar pattern as the asymmetry index in that 
the peculiar and spiral galaxies have on average higher clumpiness values
in the $z_{850}$-band than in the $H_{160}$-band. The ellipticals have on average the same
clumpiness, which is often zero, within the $H_{160}$-band and the $z_{850}$-band.

Overall, what we find is that at $z > 1$ there is less of a morphological
$k$-correction than what is seen in the nearby universe, or at least
at $z < 1$.  This was originally noted in the lower resolution imaging of
the Hubble Deep Field North when the NICMOS and WFC2 data were compared
(e.g., Dickinson et al. 2000; Conselice et al. 2005).  The meaning of
this is fairly clear - the stellar populations that make up the light
originating in the rest-frame optical is dominated by younger stars, or
at least stellar populations that contain enough young stars to still
be visible in the ultraviolet.  This implies that we are not likely to 
learn much beyond our current understanding of this aspect of galaxy
morphology in larger surveys such as CANDELS -- however, this does lead 
to the possibility of utilising the morphological $k$-correction to find
more evolved stellar population or ultra-dusty galaxies at these redshifts.

\begin{figure*}
%\vspace{5.5cm}
 \vbox to 110mm{
\includegraphics[angle=0, width=164mm]{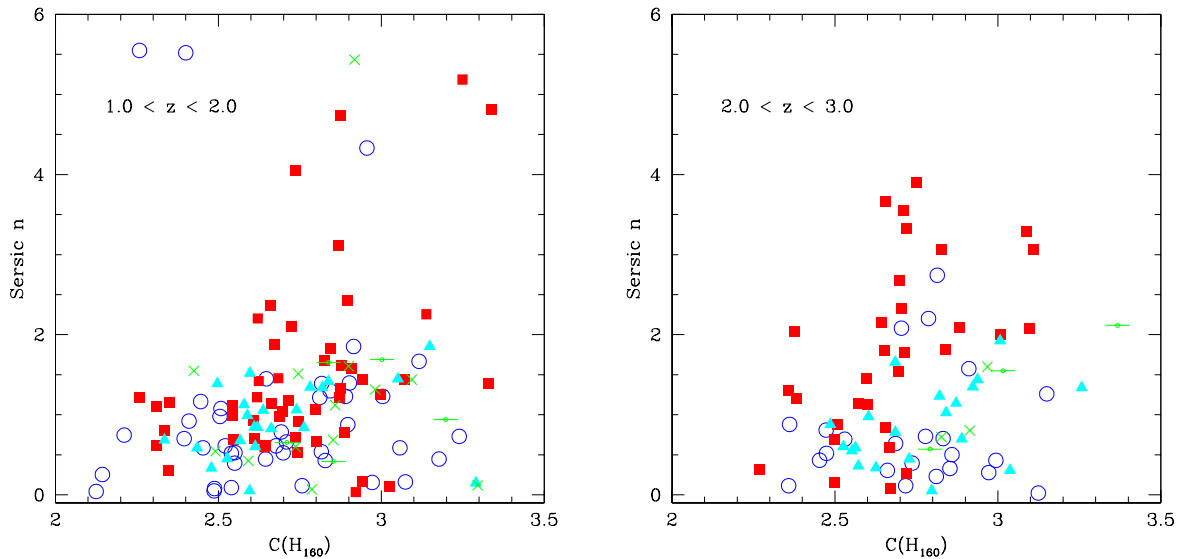}
%\vspace{-0cm}
 \caption{Plot showing the relationship between the concentration parameter $C$ and the S{\'e}rsic
index, $n$ for galaxies divided into redshift ranges $1 < z < 2$ and $2 < z < 3$.  Show on this
panel as well are the visual classification types, as in Figure~1, such
that the open blue circles are peculiars, the red boxes are the early-types,
the cyan triangles are the peculiar ellipticals, the green lines are edge-on
disks and the green crosses are disks/spirals. }
} \label{sample-figure}
\end{figure*}
%\vspace{4cm}

\subsection{Galaxy Structure at $z > 1$}

With WFC3 we are able to examine 
the rest-frame optical morphologies of galaxies at $z > 1$ in
detail using the WFC3 J$_{110}$ and H$_{160}$ imaging.  We use multiple 
approaches for this, including visual
estimates of morphological types, as well as examining the CAS
parameters and S{\'e}rsic fits to our galaxies.

The first observation we discuss is the distribution of galaxy
visual structure with morphology and stellar mass, as shown in
Figure~2. This Figure shows how our $z > 1$ sample is classified
in both the ACS and WFC3 imaging of the UDF.  The differences in these
classifications can be visualised by comparing the two plots
based on ACS and WFC3 imaging.   A discussion of
the visual apparent morphologies for these systems is included
in Conselice et al. (2008).

\begin{figure*}
%\vspace{5.5cm}
 \vbox to 105mm{
\includegraphics[angle=0, width=164mm]{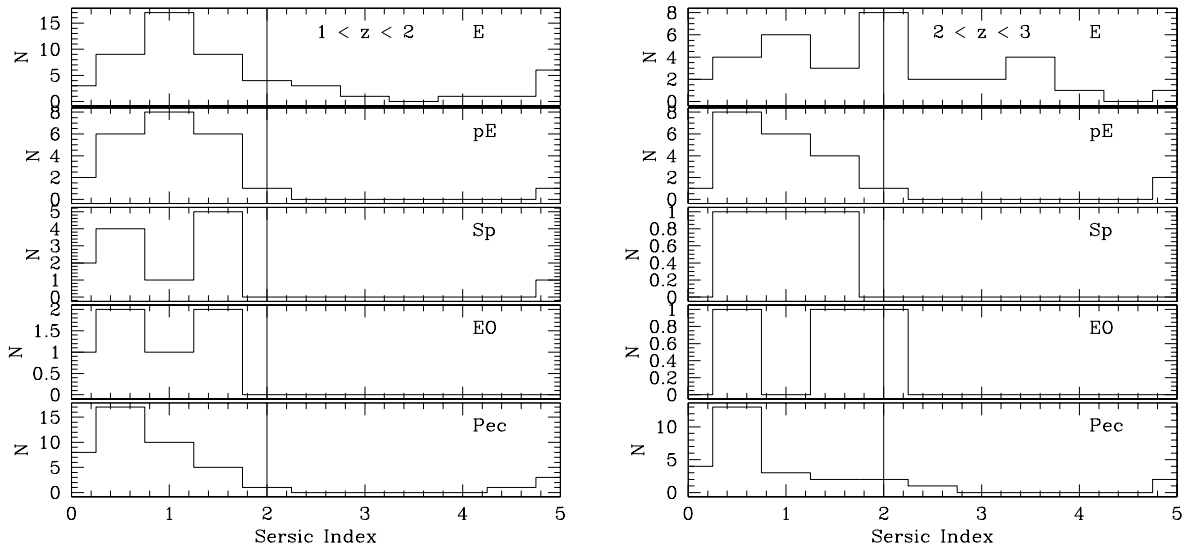}
%\vspace{-0cm}
 \caption{Figure showing the distribution of GALFIT S{\'e}rsic indices for our 
visually classified galaxies within the WFC3 UDF.  Show are two panels divided 
by redshifts, with the left hand panel galaxies in our sample found at 
$1 < z < 2$, while the right panel shows galaxies at $2 < z < 3$.  For
each visually determined galaxy type: elliptical (E), peculiar elliptical 
(pE), spiral/disk (Sp), edge-on disk (EO) and peculiar (pec), we show the 
distribution of S{\'e}rsic indices.  The vertical lines at $n = 2$ demonstrates 
the separation between galaxies that are `disk-like' with $n < 2$ and those 
that are `elliptical-like' with $n > 2$ (see Ravindranath et al. 2006; 
Buitrago et al. 2008).}
} \label{sample-figure}
\end{figure*}
%\vspace{4cm}

Figure~2 shows that there is a broad distribution of types over
the entire stellar mass and redshift range of our sample of galaxies.  The
most obvious difference between the ACS and the WFC3 classifications
is that there are fewer peculiar galaxies in the WFC3 identifications
than are seen in the ACS images.  The peculiar systems seen in the ACS 
imaging mostly become peculiar ellipticals and E/S0/Compact systems when viewed
in the rest-frame optical with WFC3, as discussed in \S 4.1.

We quantify this change by examining the distribution of morphological types 
as a function of redshift, as plotted in Figure~5.  This shows the remarkable
result that the peculiars which dominate the galaxy population within
the ACS imaging, which probes the rest-frame UV at $z > 1$, drop by
a factor of 1.5-2 in relative distribution at $z > 1$ when imaged
within the WFC3 H$_{160}$-band.   A large fraction of this change
in types goes into the elliptical and compact populations. We do not see
a large increase in the number of disk galaxies at high redshifts,
however, we do see more of these systems in the WFC3 imaging 
(see \S 4.1).

We further plot the breakdown within the elliptical classification in
Figure~6.  Roughly half of the `elliptical' population shown in Figure~5
is in the form of pure ellipticals, i.e., systems with no peculiarity.
A further roughly 20\% of the overall class of `ellipticals' are compact
and the remaining systems are peculiar ellipticals.  There is no strong
trend within the `elliptical' class within these subtypes, although the
peculiar ellipticals do appear to increase slightly at lower redshifts.

If indeed the structures of the bulk of
these galaxies are elliptical at high redshift this has important
implications for galaxy formation studies.  However, one major
issue we have to deal with is the fact that the PSF of the WFC3
is larger than ACS and therefore it is important to quantify the
effect of this larger WFC3 PSF on the classification of these galaxies.
To address this we carry out several simulations to demonstrate that
our results are robust. In \S 4.2.1 we convolve the PSF of WFC3 with
the UDF ACS image and reclassifying the same galaxies.  We also determine
our classification biases through simulating lower redshift galaxies to how 
they would appear at higher redshift in \S 4.3.

\subsection{Resolution and Depth Effect Biases}

\subsubsection{Effects of the WFC3 PSF}

In all investigations of the structures and morphologies of distant
galaxies a general problem is understanding how distant
galaxies appear, and how they would look if we could image them as we do
nearby galaxies.   We address this issue in some detail in \S 4.3.2.
Before examining this problem in depth, we first ask the simple question - 
is what we see in terms of the difference in structure between our galaxy
images in the ACS and WFC3 
due to the different PSFs, or a real effect?  

The difference between the ACS
and WFC3 imaging morphologies seen in \S 4.2 is not simply just a matter 
of wavelength differences, but
also PSF and level of depth.    In principle, any of these three can create
morphological differences (Conselice et al. 2008), and while the first one 
due to wavelength is
a `real' (i.e., not instrument driven) effect, if there are differences due 
to PSF or depth, then this is
simply due to the way the data was taken. To address this issue, we convolve 
the ACS image of the UDF
in the $z_{850}$-band where our original work was carried out 
(Conselice et al. 2008) with the WFC3 PSF
to determine how the morphologies of galaxies changes solely due to the slighter
worse resolution of WFC3.

To carry this out we reclassify our galaxies seen in the new convolved image
in the same way we did on the original ACS and WFC3 images (Figure~5). 
We carry out this classification in the same way we did our original classifications.
This allows us to determine, and separate, the contribution to the change
in morphology as resulting from different PSFs between the ACS and WFC3 camera
as opposed to differences produced from probing different rest-frame
wavelengths.

\begin{figure*}
%\vspace{5.5cm}
 \vbox to 110mm{
\includegraphics[angle=0, width=164mm]{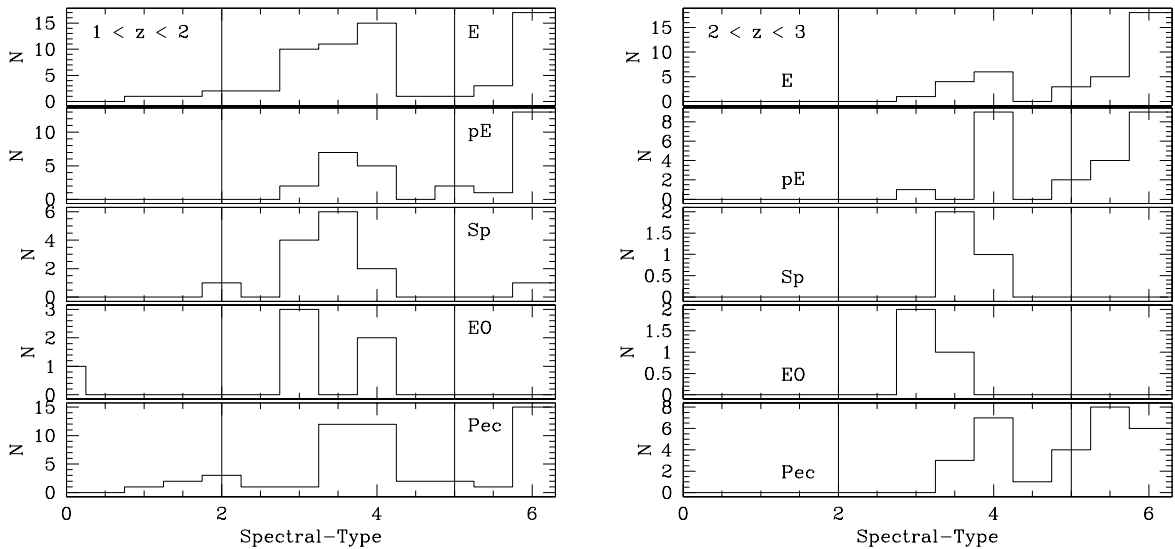}
%\vspace{-0cm}
 \caption{Figure showing the distribution of spectral types for our visually classified galaxies
within the UDF.  Show are two panels divided by redshifts, with the left hand panel galaxies
in our sample found at $1 < z < 2$, while the right panel shows galaxies at $2 < z < 3$.  For
each visually determined galaxy type: elliptical (E), peculiar elliptical (pE), spiral/disk
(Sp), edge-on disk (EO) and peculiar (pec), we show the distribution of spectral types for
each.  These types are such that: 1 = E, 2=Sbc, 3=Scd, 4=Im and 6=Starburst.  The vertical
lines show the division between early types at $< 2$, mid-types or disk at $2 - 5$, and
star forming galaxies, such as irregulars at $>5$. }
} \label{sample-figure}
\end{figure*}
%\vspace{4cm}

What we find is that the disk galaxy fraction is similar between the ACS,
WFC3,
and the ACS convolved with the WFC3 PSF image, classifications. However, there is 
clearly a difference between the fractions of peculiars and early-types when 
comparing the
ACS WFC3 PSF convolved image with the WFC3 one.  While we find a large
fraction of E/Compact systems within the WFC3 images at all redshifts, we do
not see the same results in the ACS WFC3 convolved image.  In fact, while the 
number of early-type systems in the convolved ACS imaging increases from roughly 
a 30\% fraction 
to about
45\% after convolving with the WFC3 PSF.  The peculiar fraction also drops
from roughly 60\% to 50\%.  This is significantly different from the WFC3 imaging
classifications where ellipticals make up 60\% of the population and peculiars
are at $\sim$ 30\%.  This demonstration shows that over half of our classifications 
in the WFC3 imaging are either due to identifying the correct galaxy type, or are
due to morphological $k$-corrections.

\subsubsection{Galaxy Structure Simulations}

In  the previous section we examine how galaxies in the UDF ACS image would appear after 
convolved with the WFC3 PSF.  This approach is useful when comparing the WFC3
and ACS classifications, but they do not account entirely for the effects of
redshift on the appearance of galaxies as they change through redshift. To
account for this, we carry out a series of full simulations of taking galaxies
at $z \sim 0$ to how they would appear at higher redshifts.  Within
this discussion, we use simulations of nearby galaxies placed at $z = 2.5$ as a fiducial
amount of change for nearby galaxies to within the redshifts of this study 
($1 < z < 3$).

These simulations are based on the method of simulating nearby galaxies at higher
redshift as outlined in Conselice et al. (2003). Similar simulations have 
been carried out when examining morphologies in ACS imaging in Conselice (2003) 
and Conselice et al. (2008), and for NICMOS and WFC2 imaging in 
Conselice et al. (2003).  These
simulations are also used to determine how quantitative structural parameters
change with redshift as outlined in Conselice et al. (2003, 2008).

These simulations are based on real images, taken in optical B-band, for nearby
galaxies, using the sample from Frei et al. (1996).  This sample is discussed
in terms of morphology in Conselice et al. (2000a) and Bershady
et al. (2000). Suffice it to say that this sample consists of bright nearby
galaxies, mostly ellipticals and spirals. We do not consider merger simulations
in this paper, although they have been discussed in detail in Conselice (2003),
and will be the focus of future papers on this topic.

The simulations are done essentially by taking these nearby galaxies, as imaged
with the Lowell 1.5 m telescope and placing them as they would appear in our
WFC3 imaging of the UDF.  All aspects of the Hubble Space Telescope, and the
WFC3 imager, are considered in this simulation, including its: aperture, band
width of different filters, throughput, pixel scale, read-noise, dark current,
as well as using the empirically derived sky background (e.g., Giavalisco
et al. 1996; Conselice 2003).  Furthermore, we also
experimented with two PSFs - one calculated from Tiny Tim and another based on 
stacking real stars from WFC3 imaging in other fields, such as the GOODS Early
Release Observations.

The process in carrying out these simulations is outlined in detail in 
Conselice (2003) and we
only give a summary of this here.  The nearby galaxy image is first
reduced in angular size to how it would appear at the simulated redshift. Next
the flux is reduced by  $(1+z)^{4}$ to account for surface brightness
dimming.   A background is then created which include all noise contributions, the
major one being from the WFC3 background, which is at a level of 
$\sim 0.5$ e$^{-}$ s$^{-1}$ pix$^{-1}$, which is around a factor of ten higher
than for ACS.  The galaxy is then placed into this simulated background, and 
the WFC3 PSF is applied to the image.  We increase the surface brightness of
each image by 1 magnitude, as it is known that distant galaxies are
brighter than nearby galaxies by roughly this amount, with passive evolution
with an exponentially declining star formation rate producing at least this
much evolution.

Figure~7-8 shows examples of these simulations, for galaxies both before and after
the simulation.  We only
show in this paper the disk galaxies, as mergers and ellipticals nearly always are
identifiable as such when simulated at higher redshift.  We divide these disks in early-types (Sa
through Sb; Figure~7) and late-types (Sc and Sd; Figure~8).  The results are at times 
worrying
for the simple reason that many of these galaxies would appear to be classified
visually as ellipticals. In fact, the problem is more acute for the early type disks
where the bulge component tends to dominate the appearance of the galaxy after it has
been simulated.  Viewing these galaxies as we did with {\em ds9} it would be difficult
to always identify these as disks, especially as morphological classification is a 
subjective process, and different classifiers will classify the same galaxy in a 
different way.

In general for the late-type (Figure~8) and to a lesser extent the early-type disks 
(Figure~7) the simulated galaxies appear to have
a morphology that is mostly bulge. However, these systems often have fainter outer 
`halo'-like material which can be interpreted, if identified correctly, as a disk system.
However, it is possible that these systems could be identified as ellipticals with
an outer envelope.  In this paper, we were therefore careful to carry out our classifications
of disks based on whether or not a roughly symmetrical outer portion of light was detected
in an otherwise bulge system. Regardless, it is obvious that even with deep UDF WFC3
imaging that much of the structure in spiral galaxies is lost. Clearly spiral arms can no
longer be identified, and often the disk itself is much less prominent than the bulge,
undoubtedly this is due to the lower surface brightness of disks.  This will
be a major issue when interpreting shallower CANDELS data.

We classify our simulated sample as we did our original, and find that within these conditions
we would have misclassified 1/4th of the disk galaxies in the Frei samples as `early-types'.
This is similar to the fraction of galaxies we misclassify in the ACS imaging due to the larger
PSF. The reasons for these is likely similar, and due to the larger PSF of WFC3. We account for
this in our analysis of the overall evolution of types in \S 4.4.

A major caveat to these simulations is that it is likely that disk galaxies in the early
universe are potentially different from modern disks.  We know that galaxies, even
those with a disk-like morphologies and S{\'e}rsic indices indicating disks, are smaller
by a factor of a few at $z \sim 2.5$ for at least the most massive systems (Buitrago et al. 2008).  
This will make it harder to identify disks if they are indeed smaller and therefore
more difficult to resolve. On the
other hand, we know that the quantitative morphological distribution for massive galaxies
is quite different at these redshifts than today, and that any disks at these redshifts
will likely not have as prominent bulges as contemporary disks. Furthermore, disks at high
redshifts are potentially undergoing significant amounts of star formation, making their
disks easier to identify and study.  All of these features will have to be examined in more
detail in future papers.

\subsection{Overall Results of Apparent Morphological Evolution}

In summary, the correction for PSF effects within WFC3 is at a
level of 30-40\% at most, as described in \S 4.3.1.  This is  verified by our analysis
of simulations of nearby galaxies placed at
high redshift (\S 4.3.2). In these simulations, we still find that `early-type' morphologies 
are the dominate one
at $1 < z < 3$.   At most a correction would bring down the E/compact
line in Figure~5 to roughly 40-50\% -- which is still much higher than what is implied
with the ACS imaging, as well as different from previous results (Conselice et al. 2008).  
Therefore one major
conclusion from this work is that high redshift galaxies appear to be more early-type
than disk-like.   However, this is only for our mass range, which is from roughly
log M$_{*}$ = 9 to just above log M$_{*} = 10$.    This trend is potentially different
for galaxies with log M$_{*} > 11$ as seen in the GOODS NICMOS Survey (Buitrago, Trujillo
\& Conselice 2011).  

However, as we show in the remainder of this paper, these galaxies, while appearing to be 
early-types are anything but analogs of early-type, in terms of their stellar 
populations, and they are likely
undergoing significant evolution.  Already, this can be seen by the fact that there is
a morphological $k$-correction for WFC3 identified early-types, which appear as distorted
systems in the rest-frame UV as probed by the ACS camera.   We discuss the structures of
these early-types as well as their spectral types, demonstrating the ongoing evolution of the
Hubble sequence at $1 < z < 3$.

\subsection{Quantitative Structure and Spectral Types}

\subsubsection{CAS and S{\'e}rsic Classifications}

In Figure~9 we show the locations of our sample in concentration-asymmetry space and label our
galaxies according to both their spectral and visual classification type.  This plot is
best understood by looking for boxes that agree, or otherwise, in terms of the two colours
plotted at each point.  A blue box is for a starburst/Im spectral type, and a red
box is for an E spectral type, while a green central dot is for Sbc/Scd spectral types. If the two
colours are the same then the two ways of classifying that galaxy through visual and
SED spectral-types are the same. If they differ
then they do not agree.

The most remarkable feature of this diagram is that there is a significant amount of systems
which are classified as having different spectral and morphological types throughout.  
What can be seen however, is that there
is some relation between spectral galaxy type and position on the concentration-asymmetry plane.
The starburst/Im classifications are nearly all in the higher asymmetry part of the distribution
at $z < 2$.  This however, changes at $z > 2$ where the starburst/Im types are found throughout
the concentration/asymmetry plane. This is another indication that the quantitative measurements
of structure at $z > 2$, particularly for lower asymmetry systems, cannot be interpreted in
exactly the same way as a lower redshifts (e.g., Conselice et al. 2008).  

Furthermore, as Figure~9 shows, there is not always a good agreement between visual estimates
of morphology and their positions in the CA space.    The disk galaxies, both those classified
as such visually and through spectral energy fits to SEDs are often different objects.  
Furthermore the peculiar galaxies
have a high asymmetry at each redshift, and often the peculiars are the majority of the
systems in the merger region of CAS space, with $A > 0.35$.  However, those galaxies
classified visually as early-types are often found in the region of CA space occupied
by the location of late-type galaxies in the nearby universe. This does 
not imply that
these systems are late-types, but that they have asymmetry values that are larger, with
around $<A> \sim 0.2$. This is significantly higher than the average asymmetry for early
types in the nearby universe which have values $<A> = 0.02\pm0.02$, which is more than
10 $\sigma$ different from higher redshift galaxies. This clearly shows that the visually
identified ellipticals are in a formation mode.

We furthermore show on Figure~10, the asymmetry-clumpiness plane of the CAS space.  Many objects
are too smoothed by the WFC3 PSF to measure this parameter (see Conselice 2003). However, those 
systems which are the most clumpy, as compared to their asymmetry, are the galaxies identified
visually as disk systems.  This is due to the fact that these visually identified disks are
undergoing intense star formation in clumps, and potentially, using a comparison of asymmetry
vs. clumpiness is a powerful way to identify these systems in WFC3 data.

We can get a further idea regarding the distribution of the properties of various
high redshift Hubble types by examining the S{\'e}rsic fitted $n$ index vs. the 
concentration parameter for our systems as labeled by spectral type.  Figure~11 shows this,  
where we 
plot the spectral type as different symbols.   In general, we find that galaxies with higher
S{\'e}rsic indices have higher concentration values, with most systems at $n > 2$ having a 
concentration
value $C > 2.7$, typically the value for early-types in local universe (Conselice 2003). However,
clearly there is significant scatter in these values, and the galaxies identified visually as
early-types (red boxes) are in every location of this diagram. The disk galaxy systems (the
green crosses) however show that at least the disk galaxy systems are found at lower
S{\'e}rsic indices of $n < 2$.
  
Figure~12 shows the distribution of the S{\'e}rsic indices for our sample, showing the 
range of values for each visually classified
type, and that most visually classified ellipticals have lower S{\'e}rsic indices.
This demonstrates that at least a fraction of
the early-types, or the progenitors of early-types, have the expected S{\'e}rsic morphology, 
but still have
enough residual star formation to have spectral types such as Scd or starburst (Figure~13; \S 4.5.2).  
Figure~14 on the right hand side, furthermore shows that much of the $n >2 $ systems
have a starburst spectral-type, and thus we may be seeing the epoch of the formation of
the bulk of early types at $z > 2$ when finding bona-fide early types as a non-star forming,
morphologically identifiable early-type, will be rare at these times (Bauer et al. 2011).

\subsubsection{Spectral Type Classifications}

When examining galaxies and classifying them into different types, there are several 
ways to proceed. The traditional way is to examine systems in optical light and to 
classify the galaxy onto the
Hubble sequence or a variation of the Hubble sequence.  For nearby galaxies there is a good
correlation between the Hubble type and physical parameters of the galaxy, such as the
colour, star formation history, size, etc. (e.g., Roberts \& Haynes 1994; Conselice 2006a).  
However, this relationship may
not hold at higher redshifts.  Therefore, it is desirable to classify galaxies into
spectral types, as well as visual estimates of morphology. 

Spectral types can be obtained through high S/N spectra, or through fitting of the spectral
energy distributions of the galaxies.  The approach that we use in this paper is
to fit spectral energy distributions to the broad-band photometry of our sources and to
derive which is the best fitting spectral type - elliptical, early-type spirals, late-type
spirals, irregulars and starbursts (\S 3.4) .   This is similar to our morphological scheme in
principle, although there is no reason why these spectral-types should match morphological
ones at $1 < z < 3$  as well as they do at lower redshifts.

First, we give some numbers on the overlap of these types.  At $1 < z < 2$, we find that
for systems identified as disk galaxies and edge-on disk galaxies visually,  86\% (83\%) of
systems have a spectral-type consistent with a disk, i.e., Sbc/Scd types 
(where the second number is parenthesis
is the edge-on disk fraction).  For the remainder, we find that 7\% (17\%) of disks have an
early-type spectral-type,  and 7\% (0\%) have spectral types that are star forming/starbursts.
What this shows is that our visual estimates are roughly correct in identifying which
systems have a mix of stellar populations - i.e., a mixture of an old and a younger component.
At higher redshifts ($2 < z < 3$) we find for both types of disks that 100\% are 
identified as Sbc/Scd spectral types.  

For systems identified as visual early-types in WFC3 imaging, we find a more mixed outcome.  
At $1 < z < 2$ 
we find that only 10\% of these systems would be identified as E-spectral types. That is, 
only a very small
fraction of the WFC3 imaged UDF galaxies we find visually as early-types have a spectral 
shape suggesting that
they contain evolved stellar populations.    The bulk of the visually classified early-types are
found to be consistent with a Sbc/Scd spectra type (56\%) at $1 < z < 2$, suggesting a mixture
of old+young stellar populations, or the fading of an older stellar population that did not
form in one complete burst.  Interestingly, 34\% of the visually classified early-types 
at $1 < z < 2$ are consistent with being starbursts or Im systems - that is galaxies 
which are dominated by star formation.  This rises to 62\% starburst spectral-types
for visually classified 
early-types at $2 < z < 3$.    We will return to a discussion of this issue in \S 5 
and what it implies for early-type formation.

The two final visual categories we discuss are the peculiars and the peculiar early-types, which
are systems that resemble early-types in some formation state and display a visual peculiarity.
What we find is that at the higher redshift range, $2 < z < 3$, these two types
have similar spectral energy distributions -- the visually classified peculiars 
are 62\% starburst/Im spectral type, and 38\% Sbc/Scd type. Very interestingly, this is just
slightly less than the fraction of visually classified early-types which are star forming.
As for the peculiar early-types, we find a very similar fraction of 60\% Im/starburst spectral-type and
40\% Sbc/Scd.  Based on this, it appears that the peculiar/E/pE types may have a similar
origin, and this is reflected in the similar spectral types of all these types.

\begin{figure*}
%\vspace{5.5cm}
 \vbox to 110mm{
\includegraphics[angle=0, width=164mm]{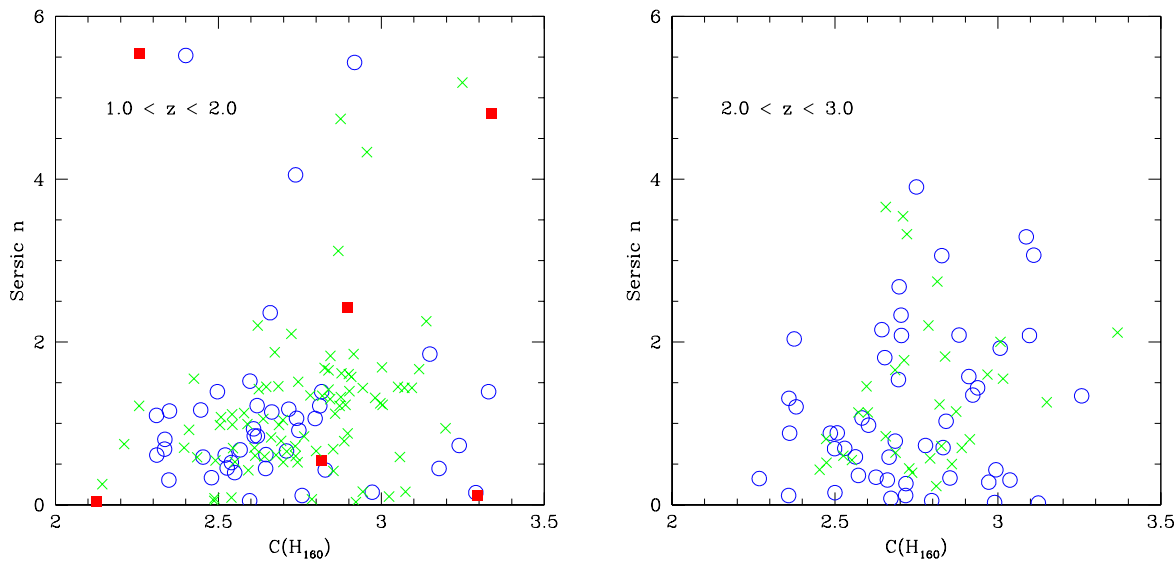}
%\vspace{-0cm}
 \caption{Plot showing the relationship between the concentration parameter $C$ and the S{\'e}rsic
index, $n$ for galaxies divided into redshift ranges $1 < z < 2$ and $2 < z < 3$.  Show on this
panel as well are the spectral classification types, with the following labelling scheme:
the solid red open boxes are classified as early-type spectrally,
open blue circles are starforming systems, the Im/starbursts, and the green crosses are 
those classified spectrally as disk like, or Sbc/Scd.}
} \label{sample-figure}
\end{figure*}
%\vspace{4cm}

At lower redshifts, $1 < z < 2$, we find that the fraction of peculiars which remain starburst/Im
systems drops to 35\%, with 10\% having an E spectral type, and 55\% having a Sbc/Scd spectral
type. This is very similar to the breakdown for the ellipticals.
This shows that peculiar galaxies are not all starbursting systems, but likely
are distorted due to perhaps the merging of existing galaxies which contain a mixture of old and
young stellar populations.  The peculiar ellipticals at $1 < z < 2$ are 53\% starburst/Im
spectral types and 47\% Sbc/Scd spectral types, showing a similar pattern to the peculiars, 
but with even more systems having a starburst origin. 

\begin{figure*}
%\vspace{5.5cm}
 \vbox to 90mm{
\includegraphics[angle=0, width=184mm]{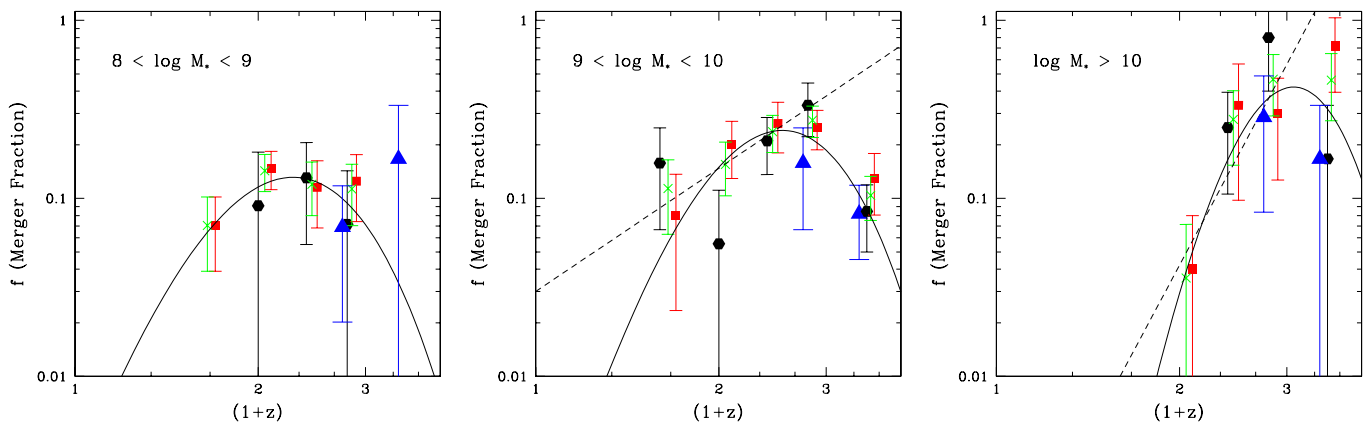}
%\vspace{-0cm}
 \caption{The merger fraction as a function of redshift and stellar
mass for our WFC3 sample (blue triangles) compared with data from the Hubble-Deep Field North
NICMOS (red squares) and the ACS Ultra-Deep Field (black circles).   
Shown in this figure are the merger fractions at mass limits of $8 <$ log M$_{*} < 9$,
$9 <$ log M$_{*} < 10$, and log M$_{*} >$ 10.  The green crosses show
the evolution of the merger fraction for a combined UDF and HDF-N sample.}
} \label{sample-figure}
\end{figure*}
%\vspace{4cm}

\subsection{Tracing Galaxy Formation and Actual Types}

There are two goals to studying galaxies in terms of their structure in this paper. 
These are  
understanding when and how the Hubble sequence was put into place. Clearly, as we have
just shown, the Hubble sequence, as we know it today, does not exist at higher
redshifts, particularly at $z > 2$. The question becomes - how can we identify galaxies
at higher redshifts that match the properties of nearby Hubble types? This goes beyond
morphology and asks the question - when do ellipticals, in the sense of smooth, nearly
structureless galaxies dominated by old stellar populations, first form?  Related to this
is the formation of galaxies with disks.  The former problem of the ellipticals is easier
to probe, in principle, as finding a galaxy with a smooth morphology and older stellar populations
is more straightforward than determining if a galaxy is a rotating disk.  We also investigate whether
we can use structure to determine the formation modes of galaxies, where we consider the merger
history in \S 4.6.3.

\subsubsection{A `Pure' Elliptical Evolution}

First, we discuss the fraction of pure ellipticals within the UDF.  We define pure 
ellipticals as galaxies which have a spectral-type consistent with an early-type, and a
morphology consistent with being a modern elliptical, which we select as
$A < 0.2$.   At $z = 0$ the elliptical fraction for galaxies with stellar
masses $M_{*} > 10^{10}$ \solm is $> 20$\% (Conselice 2006a).  With our selection,
we find a much smaller fraction at higher redshifts.

When we examine the evolution of this pure elliptical fraction with
redshift we find that there are no pure early-types at $z > 2$.  We find a total
of six systems at $1 < z < 2$ giving a fraction of 3.2$\pm$2.3\% at $1.5 < z < 2.0$
and 4$\pm$2\% at $1 < z < 1.5$.  We find a higher fraction if we limit our
selection to $M_{*} > 10^{10}$ \solm systems =  14$\pm$14\% at $1.5 < z < 2$ and
23$\pm$13\% at $1 < z < 1.5$.  This demonstrates that massive galaxies are more 
likely to be elliptical than lower mass ones, which is also seen in the
nearby universe (e.g., Conselice 2006a).  This also shows that at least within
the very small field of view of the UDF, there are no pure elliptical
galaxies at $z > 2$.  However, most certainly progenitors of pure ellipticals
exist at these redshifts, and the fact that there are massive galaxies, even
within our own sample, at $z > 2$ shows that these galaxies are in an active
assembly mode, and can be identified and studied.  Future studies of larger fields,
especially within the CANDELS programme, will address these issues, and
put better constraints on the number of passive ellipticals at $z > 2$

\subsubsection{The Disk Galaxies}

Tracing disk galaxies in a similar way as the ellipticals is not as straightforward.
However, the disk galaxies at the magnitude and stellar mass limits we use are extremely
common in the nearby universe, with a fraction of 60-75\%.  Thus, just from this, a
significant number of the distant galaxies we see must be progenitors, in some form, of
disk galaxies in today's universe.
The issue with disks is that the spectral-type is a mixed population of older
and younger stars, which is not an unusual situation at high redshift, and does not
necessarily imply that these systems are always disks - consisting of an older
bulge with a younger disk.  Structurally disks can be identified by having a modest
asymmetry and concentration.  When we examine systems within our
sample with $C < 3$ and $0.2 < A < 0.35$, and spectral types consistent with
nearby disks, we find a roughly even fraction of disks of 7-15\%  between
redshifts of $1 < z < 3$, which is similar to the visual fraction (Figure~5).

However, when we use these limits we find that the galaxies we select
are not always identifiable as disks visually, in fact they often are not.  We
in fact classify most of these systems as peculiars or compact galaxies, showing that
the apparent morphology of galaxies can be quite different from what their structure
and stellar populations are revealing.

An alternative method for understanding the limits of the number of disks we have
is by examining the fraction of systems with large inclinations, which are likely disks
seen edge-on.  Through the use of our measured ellipticities with GALFIT, we find that 
roughly 30\% of our selected systems have an axis ratio b/a $< 0.3$.  If all of these 
are edge on systems, then every
galaxy at these redshifts from $z = 1 - 3$ would have to be a `disk' galaxies. While this
is obviously unlikely the case, it shows that we are likely misidentifying some 
disks, although many of these disks are likely in some formation mode and cannot
be identified as such. This issue was furthermore examined recently for
massive compact quiescent galaxies at $z \sim 2$ by van der Wel et al. (2011) who found 
that 65$\pm$15\% of massive compact systems are likely disks in some form.   While it is
unlikely that every galaxies in our sample is a disk in some form, it does show that
we are very likely under representing the number of disk galaxies in our sample by using
either visual morphologies or parametric parameters (see also Buitrago et al. 2011, in prep).

\subsubsection{The Merger History}

In this section we revisit the measurement of the merger history which
was used in Conselice et al. (2008) with ACS UDF data, and Conselice et al. (2003) using NICMOS
imaging of the HDF-N to trace the merger history for
galaxies as a function of stellar mass at $z > 1$.  Both of these studies found similar results,
that the merger history is highest for the most massive galaxies at $z \sim 2.5$, and declines
steeply at lower redshifts, while lower mass systems have a more gradual drop off in 
their merger fraction.  
Conselice et al. (2008) and Conselice (2006b) investigate, using these merger fractions, the
likely merger rate, and the contribution of merging to building up the stellar
masses of galaxies over time.

We use WFC3 data from this paper to make a new measurement of the merger
history with WFC3 data at $z > 1$, and compare this with previous work.  The errors on 
these measurements
are large simply because of the small number of galaxies used in this study. Future
work using data such as CANDELS will address the merger history issue in much more detail.  
For now however, we simply describe the merger history, and how it agrees with previous results.

First we describe our measured merger fractions based on the CAS parameters. This is a topic
with a large history and background, but the most relevant papers for measuring the merger
history with CAS are Conselice (2003), Conselice et al. (2003), Conselice et al. (2008) and
Conselice, Yang \& Bluck (2009).   As described in Conselice et al. (2000a,b) and Conselice (2003) 
one method for finding mergers is to use the conditions in rest-frame optical light (cf. Taylor
et al. 2007; Lanyon-Foster et al. 2011 for UV),

\begin{equation}
A > 0.35\, \&\, A > S,
\end{equation}

\noindent such that the asymmetry is higher than $A = 0.35$, and the asymmetry value is 
higher than the clumpiness value, $A > S$.  This ensures not only that a galaxy is 
roughly 1/3 asymmetric in terms of
its light distribution, but also that this light does not come from clumpy light, 
such as star forming regions.  This method works well at lower redshifts (Conselice 2003), yet with
the larger star forming clumps seen within high-$z$ galaxies (e.g., Elmegreen et al. 2005) it 
remains possible that some galaxies with large star forming complexes are being misidentified 
as a merger.

We show the merger fraction using our new WFC3 UDF imaging in Figure~15, 
finding a very
similar merger fraction as before using ACS.  One way to quantify this merger history is through various
fits of the merger fraction history.   
The traditional power-law format (Conselice et al. 2003a,b; Bridge et al. 2007; Conselice
et al. 2009; Bluck et al. 2009) is given by:

\begin{equation}
f_{\rm m}(z) = f_{0} \times (1+z)^{m}
\end{equation}

\noindent where $f_{\rm m}(z)$ is the merger fraction at a given
redshift, $f_{0}$ is the merger fraction at $z = 0$, and
$m$ is the power-law index for characterising the merger
fraction evolution.  An alternative way to characterise the 
merger fraction evolution is the Press-Schechter formalism 
(Conselice et al. 2008) - which is essentially a 
combined power-law+exponential form.  However, we do not consider
this form in this paper, but will in future investigations of the
merger history.  

By fitting equation 6, we find a similar
parameterisation of the merger history as we found
for galaxies with an $m$ exponent of $m \sim 5.5$, for
massive galaxies and $m \sim 2.2$ for lower mass systems.
We also show on Figure~15 the best fitting exponential/power-law
form for the merger history.  Integrating the merger rate, using
a time-scale of 0.4 Gyr (Conselice et al. 2008) we find that 
on average there are 4.3$^{+0.8}_{-0.8}$ major mergers at $ z < 3$
for the most massive galaxies with $M_{*} >$ \mass.  This is enough
merging to roughly double the mass of these galaxies (Conselice 2006b).
In the future, we will utilising more wide-field imaging from CANDELS
and other WFC3 surveys to obtain a more accurate and definite merger
history for these galaxies.

\subsection{Comparison to Models}

In this section we carry out a comparison between our results, and models of
galaxy formation which can predict the evolution of galaxy morphology as
a function of redshift.  The models we compare with are the {\em Galacticus}
semi-analytical
models from Benson \& Devereux (2010) and Benson (2010)
which do not resolve the morphologies 
of galaxies as some numerical
models do. However, these models follow the formation of galaxies
in a cosmological context, and can be used to predict how spirals, ellipticals
and galaxy mergers evolve through time based on the physics which is thought
to form these systems.\footnote{The exact model we use is version 0.9.0.r249 of
Galacticus.}

Currently, the idea is that galaxies are formed from mergers and accretion.  
Bulges and ellipticals are formed in merger events that have dynamically cooled and
relaxed, while disks and spirals are formed by the smooth accretion
of intergalactic gas which is then converted into stars over some time period.
These models predict these quantities, 
as well as the time since the last major merger for each galaxy at a given redshift.

We use these models to predict the morphological evolution of galaxies in a cosmological
context, and therefore whose formation is influenced by the relative distribution of
dark matter, dark energy and the temperature of the dark matter.  These
models are such that a measure of the bulge to total ratio can be derived based
on the history of spheroid and disk assembly.    Thus, while these
models do not resolve galaxy morphology, they can predict what the likely morphological
distribution is based on the physics of the galaxy formation.

This comparison is shown in Figure~16 between our three main types and the same 
predictions from Benson \& Devereux (2010).  The way to view this comparison is that the
data has the same line type as in Figure~16, and the predictions of the same quantities
are shown as double lines.  The green doubled dashed line shows the evolution of
the disk fraction, as defined by B/T $< 0.5$.  The red short dashed line shows predictions
for the evolution of the elliptical fraction, as defined by B/T $> 0.9$.  The
merger fraction predicted in the simulation are galaxies defined as having had
a major merger
in the past 1 Gyr, similar to what would be predicted for producing a peculiar
galaxy (Conselice 2006b).  

\begin{figure}
\hspace{-0.5cm}
 \vbox to 150mm{
\includegraphics[angle=0, width=90mm]{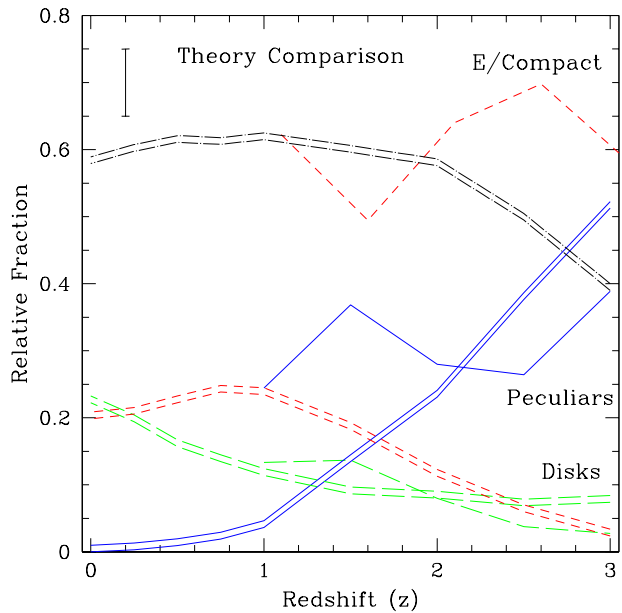}
 \caption{The comparison of our WFC3 morphologies, probing
the rest-frame optical at $1 < z < 3$ and model predictions
from Benson \& Devereux (2010).  The single lines are the data, as shown
in Figure~5, where the dashed line red line is the E/Compact
systems, the solid blue line denote the peculiar systems and
the dashed green line is for those galaxies classified as
disks.  The double dashed lines, with the same colour and
line type as the data, are the model predicts from Benson \& Devereux (2010).
In the models the E/Compact is for systems with B/T $> 0.9$, and
the disks are systems with B/T $< 0.5$, and the peculiars
are systems which have undergone a merger in the past Gyr. The
black dot-dashed line shows the fraction of systems with
$0.5 <$ B/T $<0.9$.}
%\vspace{5cm}
} \label{sample-figure}
\end{figure}

The comparison between these quantities is fairly good using these definitions, which
are somewhat arbitrary as we do not know, nor have an obvious way, to compare which
B/T ratios correlate with early/late type disks.  This is demonstrated by the white
dot-dashed, which is the fraction of systems with $0.5 <$ B/T $<0.9$, which ranges from
a fraction of roughly 0.4 to 0.6 within our range of interest.  If we consider these
systems are `early-type', or based on the results in \S 4.3 as indeed true disks 
which appear as early-types in our classification due to the effects of redshift,
then the predicted fraction would agree quite well with the early-type
fraction we observe.  However, previous work on more detailed comparisons between
mergers and theory predictions show a worse agreement (Bertone \& Conselice
2009).

\section{Summary}

This paper presents an analysis of galaxy morphology and structure at
 $1 < z < 3$ using WFC3
data utilising several popular approaches, include visual estimates of
morphology, CAS parameters, parametric S{\'e}rsic profile fitting, as well
as spectral-type fitting to spectral energy
distributions.  We furthermore analyse morphological $k$-corrections
between ACS and WFC3, essentially probing the difference in rest-frame
morphology between the ultra-violet and optical light at $z > 1$. 
Our major findings include:

\vspace{0.25cm}

\noindent I. In terms of morphological $k$-corrections, a large fraction
of systems classified as `peculiar' at $1 < z < 3$ within ACS
imaging appear as early-types in the WFC3 imaging.  This result
holds to some degree even after convolving the ACS image with the larger WFC3 
PSF, and by simulating nearby galaxies to high-$z$, and redoing 
classifications.    However, these
`early-types' are clearly not the same as those at $z = 0$ in terms of
their stellar populations, showing evidence for star formation and
ongoing assembly.

\vspace{0.25cm}

\noindent II.  We further find that these `early-type'classifications
dominate the non-peculiar galaxy population
at $1 < z < 3$, with a fraction near 30\%, with a similar fraction of
peculiars.  The remaining systems are classified as `compact' galaxies,
and disk-like galaxies, which are both on average 10-20\% of the population. 

\vspace{0.25cm}

\noindent III.  We show that there is a significant difference in the population
of galaxies classified as early/mid/late types as defined
by spectral-types and those
defined through visual estimates. The most common difference is
found for galaxies classified as elliptical -- a significant fraction of
these have a disk spectral type, or even a starburst spectral type.
This shows that very few purely passive massive galaxy exist at $1 < z < 3$,
at least within the UDF.  

\vspace{0.25cm}

\noindent IV. Examining the CAS parameters, we find that the change in CAS
values between the $z_{850}$ and $H_{160}$ bands is similar, per unit
wavelength, as found between the rest-frame UV and optical at $z < 1$
(Conselice et al. 2008).  

\vspace{0.25cm}

\noindent V.  By using a strict definition of elliptical and disk
galaxies that matches the concentration, asymmetry, and spectral
types for nearby systems, we find very few `formed' systems at
$z > 1$.  We find 10\% of our sample would be classified as
nearby disks at $z > 1$, compared with 75\% in the modern universe.
Likewise, we find $<5$\% of systems classifiable as modern
ellipticals, about a factor of three lower than today.

\vspace{0.25cm}

\noindent VI. We find that the inferred merger fraction for the most
massive galaxies ($>10^{10}$ \solm) at $z > 1$ is between 20-30\%. This is very similar
to previous findings using UDF ACS and NICMOS HDF imaging, revealing
that even when probing the rest-frame optical at $z > 1$ we find a 
similar merger history, and therefore a similar contribution of mergers towards
forming galaxies, as when looking at lower resolution and bluer wavelengths.

\vspace{0.25cm}

\noindent VII. We find that the properties of visually classified elliptical
peculiars, and the peculiar ellipticals are very similar, suggesting that
these galaxies are in some common formation scenario. It is very likely that this common
process is galaxy merging, with each type in a different phase of merging. 
We are therefore likely seeing the formation modes of
elliptical galaxies through these systems.

\vspace{0.25cm}

We do not find a large fraction of visually classified disk galaxies at $z > 1$ (10-15\%).
This is potentially due to misidentifying ellipticals as rotating disks, however the
spiral structure, and star forming regions seen in disks at $z < 1$ and at times
at $ z > 1$ should be visible within our imaging.  The number of edge-on disk
galaxies, and the high number of inclined systems we see, 
suggests that perhaps we are misidentifying some face-on disks
as ellipticals, however, the mode of formation of these systems must be quite
different from spirals in the more nearby universe.

Overall, our conclusion is that visual estimates of morphology are limited,
both because of the PSF of WFC3 limits our ability to resolve distant galaxies,
as well as the intrinsic nature of morphology at higher redshifts. We demonstrate
that the best way to find distant galaxy `type' analogs to modern galaxies, and
therefore to address the question of how and when the Hubble sequence forms is
through quantitative measurements of both the stellar populations and quantitative
structures, and leaving behind visual estimates.

\vspace{0.25cm}

We acknowledge support for this work from STFC in the form of post-doctoral research assistant and
studentship support, and the Leverhulme Foundation in the
form of a Leverhulme Trust Prize to CJC.  We thank Andrew Benson for providing us with the
latest output of his Galacticus semi-analytical model.

\label{lastpage}

\end{document}